\documentclass[11pt,a4paper,english]{article}
\usepackage[T1]{fontenc}
\usepackage[latin9]{inputenc}
\usepackage{amsmath}
\usepackage{graphicx}
\usepackage{amssymb}
\usepackage{esint}

\makeatletter

\providecommand{\tabularnewline}{\\}

\usepackage{a4wide}
\usepackage{psfrag}
\makeatletter
\renewcommand{\theequation}{\hbox{\normalsize\arabic{section}.\arabic{equation}}}
\@addtoreset{equation}{section}
\renewcommand{\thefigure}{\hbox{\normalsize\arabic{section}.\arabic{figure}}}
\@addtoreset{figure}{section}
\renewcommand{\thetable}{\hbox{\normalsize\arabic{section}.\arabic{table}}}
\@addtoreset{table}{section}
\makeatother 

\usepackage{babel}
\makeatother

\begin{document}

\title{\begin{flushright}{\normalsize ITP-Budapest Report No. 639}\end{flushright}\vspace{1cm}Finite
temperature expectation values of boundary operators}

\author{G. Takács\\
\emph{}\\
\emph{HAS Research Group for Theoretical Physics}\\
\emph{H-1117 Budapest, Pázmány Péter sétány 1/A.}}

\date{17th June 2008}

\maketitle
\begin{abstract}
A conjecture is presented for the thermal one-point function of boundary
operators in integrable boundary quantum field theories in terms of
form factors. It is expected to have applications in studying boundary
critical phenomena and boundary flows, which are relevant in the context
of condensed matter and string theory. The conjectured formula is
verified by a low-temperature expansion developed using finite size
techniques, which can also be used to evaluate higher point functions
both in the bulk and on the boundary. 
\end{abstract}

\section{Introduction}

The aim of the present work is to calculate the thermal one-point
function of local boundary operators in integrable boundary quantum
field theories. Such a theory can be specified with a Euclidean action
of the form\begin{equation}
\mathcal{A}=\int_{-\infty}^{\infty}d\tau\left(\int_{-\infty}^{0}dx\,\mathcal{L}\left(\Phi^{\alpha},\partial_{\tau}\Phi^{\alpha},\partial_{x}\Phi^{\alpha}\right)+\mathcal{L}_{B}\left(\Phi^{\alpha}(x=0),\partial_{\tau}\Phi^{\alpha}(x=0)\right)\right)\label{eq:baction}\end{equation}
where the field variables are denoted $\Phi^{\alpha}$. The bulk equations
of motion follow from the Euler-Lagrange equations specified by $\mathcal{L}$,
while the boundary condition is obtained by varying $\mathcal{L}_{B}$;
the possible choices for the action are restricted by requiring integrability
\cite{GZ}. 

\begin{figure}
\begin{centering}
\psfrag{xinf}{$-\infty\;\leftarrow\; x$}
\psfrag{R}{$R=1/T$}
\psfrag{O}{$\mathcal{O}$}
\psfrag{x=0}{$x=0$}\includegraphics{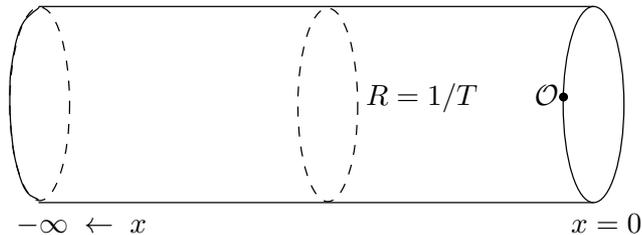}
\par\end{centering}

\caption{\label{fig:infinitecylinder} The finite temperature boundary quantum
field theory with a local boundary insertion $\mathcal{O}$}

\end{figure}

For a finite temperature $T$ the Euclidean time $\tau$ must be compactified
to a volume \[
R=\frac{1}{T}\]
Consider a local operator $\mathcal{O}$ inserted at the boundary
$x=0$ as shown in figure \ref{fig:infinitecylinder}. The quantity
of interest is the thermal average\begin{equation}
\langle\mathcal{O}\rangle^{R}=\frac{\text{Tr}\left(\mathrm{e}^{-RH}\mathcal{O}\right)}{\text{Tr}\left(\mathrm{e}^{-RH}\right)}\label{eq:thetarget}\end{equation}
where $H$ is the Hamiltonian corresponding to the action (\ref{eq:baction})
and the trace is taken on the space of states allowed by the boundary
condition. 

The main motivation to study finite temperature correlators of boundary
operators comes from boundary renormalization group flows, where the
most useful quantity characterizing the space of the flows is the
Affleck-Ludwig $g$-function or boundary entropy \cite{affleckludwig}.
The original setting where this function was introduced already made
use of finite temperature. Furthermore, as shown by Friedan and Konechny
\cite{friedankonechny}, the variation of this function along the
flow can be computed via a sum rule that is expressed in terms of
finite temperature boundary two-point functions. The present paper
can be considered as a step towards constructing such correlators
from field theory data. In addition, quantities like the thermal average
(\ref{eq:thetarget}) may have direct physical relevance to condensed
matter systems.

Our goal is to express the thermal average in terms of matrix elements
(form factors) of the operator $\mathcal{O}$. Therefore in section
2 the boundary form factor bootstrap is presented, slightly extended
from its original formulation in \cite{bffprogram} to include theories
with more than one particle species. In section 3 we formulate a conjecture
for the thermal average (\ref{eq:thetarget}) based on the earlier
work by Leclair and Mussardo \cite{leclairmussardo} in the bulk case. 

In order to provide evidence for the conjecture, the proposed formula
is developed in a low-temperature series, with the details described
in appendix A. The low-temperature expansion of (\ref{eq:thetarget})
is then evaluated using an independent method developed in \cite{fftcsa2}.
This approach requires the knowledge of boundary form factors in finite
volume (up to corrections that decay exponentially with the volume).
Section 4 presents the relevant results from the paper \cite{bfftcsa},
and appendix B provides some further details on the evaluation of
diagonal matrix elements. The calculation itself is presented in section
5, with a particularly complicated part relegated to appendix C. Section
6 is devoted to the conclusions.

\section{The boundary form factor bootstrap}

The relations satisfied by the form factors of a local boundary operator
were derived in \cite{bffprogram}. Compared to the equations in \cite{bffprogram},
the ones presented here are slightly generalized to allow for more
than one particle species. Such an extension was first given in \cite{ca2};
the derivation of these equations is straightforward using the methods
of \cite{bffprogram}. 

Here the equations are listed without much further explanation. Take
an integrable boundary quantum field theory in the (infinite volume)
domain $x<0$, with $N$ scalar particles of masses $m_{a}$ ($a=1\dots N$).
As usual in two-dimensional field theory, asymptotic particles are
labeled with their rapidities $\theta$, and their energy and momentum
reads\[
E_{a}\pm p_{a}=m_{a}\mathrm{e}^{\pm\theta_{a}}\]
Both the bulk and boundary scattering are assumed to be diagonal and
given by the two-particle $S$ matrices \begin{equation}
S_{a_{1}a_{2}}(\theta_{1}-\theta_{2})=\mathrm{e}^{i\delta_{a_{1}a_{2}}(\theta_{1}-\theta_{2})}\label{eq:tpphasehsift}\end{equation}
(where $\delta_{a_{1}a_{2}}(\theta_{1}-\theta_{2})$ are the two-particle
phase-shifts) and the one-particle reflection factors \[
R_{a}(\theta)\]
satisfying the boundary reflection factor bootstrap conditions of
Ghoshal and Zamolodchikov \cite{GZ}. For a local operator $\mathcal{O}(t)$
localized at the boundary (located at $x=0$, and parametrized by
the time coordinate $t$) the form factors are defined as\begin{eqnarray*}
 &  & \,_{a_{1}'\dots a_{m}'}\langle\theta_{1}^{'},\dots,\theta_{m}^{'}\vert\mathcal{O}(t)\vert\theta_{1},\dots,\theta_{n}\rangle_{a_{1}\dots a_{n}}=\\
 &  & \qquad F_{a_{1}'\dots a_{n}';a_{1}\dots a_{n}}^{\mathcal{O}}(\theta_{1}^{'},\dots,\theta_{m}^{'};\theta_{1},\dots,\theta_{n})e^{-imt(\sum\cosh\theta_{i}-\sum\cosh\theta_{j}^{'})}\end{eqnarray*}
using the asymptotic states introduced in \cite{BBT}. They can be
extended analytically to complex values of the rapidity variables.
With the help of the crossing relations derived in \cite{bffprogram}
all form factors can be expressed in terms of the elementary form
factors\begin{equation}
\,\langle0\vert\mathcal{O}(0)\vert\theta_{1},\dots,\theta_{n}\rangle_{in}=F_{a_{1}\dots a_{n}}^{\mathcal{O}}(\theta_{1},\dots,\theta_{n})\label{eq:elementaryff}\end{equation}
which can be shown to satisfy the following equations:

I. Permutation:

\begin{center}
\begin{eqnarray}
 &  & F_{a_{1}\dots a_{i}a_{i+1}\dots a_{n}}^{\mathcal{O}}(\theta_{1},\dots,\theta_{i},\theta_{i+1},\dots,\theta_{n})=\label{eq:permutation}\\
 &  & \qquad S_{a_{i}a_{i+1}}(\theta_{i}-\theta_{i+1})F_{a_{1}\dots a_{i+1}a_{i}\dots a_{n}}^{\mathcal{O}}(\theta_{1},\dots,\theta_{i+1},\theta_{i},\dots,\theta_{n})\nonumber \end{eqnarray}

\par\end{center}

II. Reflection:\begin{equation}
F_{a_{1}\dots a_{n}}^{\mathcal{O}}(\theta_{1},\dots,\theta_{n-1},\theta_{n})=R_{a_{n}}(\theta_{n})F_{a_{1}\dots a_{n}}^{\mathcal{O}}(\theta_{1},\dots,\theta_{n-1},-\theta_{n})\label{eq:reflection}\end{equation}

III. Crossing reflection: \begin{equation}
F_{a_{1}\dots a_{n}}^{\mathcal{O}}(\theta_{1},\theta_{2},\dots,\theta_{n})=R_{a_{1}}(i\pi-\theta_{1})F_{a_{1}\dots a_{n}}^{\mathcal{O}}(2i\pi-\theta_{1},\theta_{2},\dots,\theta_{n})\label{eq:crossing_reflection}\end{equation}

IV. Kinematical singularity\begin{eqnarray}
 &  & -i\mathop{\textrm{Res}}_{\theta=\theta^{'}}F_{aa'a_{1}\dots a_{n}}^{\mathcal{O}}(\theta+i\pi,\theta^{'},\theta_{1},\dots,\theta_{n})=\label{eq:kinematical}\\
 &  & \qquad\mathbb{C}_{aa'}\left(1-\prod_{i=1}^{n}S_{aa_{i}}(\theta-\theta_{i})S_{aa_{i}}(\theta+\theta_{i})\right)F_{a_{1}\dots a_{n}}^{\mathcal{O}}(\theta_{1},\dots,\theta_{n})\nonumber \end{eqnarray}
where $\mathbb{C}_{aa'}=\delta_{\bar{a}a'}$ is the charge conjugation
matrix ($\bar{a}$ denotes the antiparticle of species $a$).

V. Boundary kinematical singularity

\begin{equation}
-i\mathop{\textrm{Res}}_{\theta=0}F_{aa_{1}\dots a_{n}}^{\mathcal{O}}(\theta+\frac{i\pi}{2},\theta_{1},\dots,\theta_{n})=\frac{g_{a}}{2}\Bigl(1-\prod_{i=1}^{n}S_{aa_{i}}\bigl(\frac{i\pi}{2}-\theta_{i}\bigr)\Bigr)F_{a_{1}\dots a_{n}}^{\mathcal{O}}(\theta_{1},\dots,\theta_{n})\label{eq:boundary_kinematical}\end{equation}
where $g_{a}$ is the one-particle coupling to the boundary\begin{equation}
R_{a}(\theta)\sim\frac{ig_{a}^{2}}{2\theta-i\pi}\quad,\quad\theta\sim i\frac{\pi}{2}\label{eq:gdef}\end{equation}

There are also further equations corresponding to the bulk and boundary
bootstrap structure (i.e. bound state singularities of the scattering
amplitudes $S$ and $R$), but they are not needed in the sequel.
The equations are supplemented by the assumption of maximum analyticity
i.e. that the form factors only have the minimal singularity structure
consistent with the bootstrap equations. We remark that it is a general
property of non-trivially interacting diagonal factorized scattering
theories that their amplitudes are fermionic:\[
S_{aa}(0)=-1\]
and as a result of eqn. (\ref{eq:permutation}) all form factor functions
satisfy an exclusion property (Pauli principle), i.e. they vanish
when any two of their rapidity arguments coincide, together with the
corresponding species indices. 

It was shown in \cite{bffcount} that the space of solutions of the
above equations is consistent with the operator spectrum predicted
by boundary conformal field theory in the Lee-Yang and sinh-Gordon
model. More recently the author gave a general procedure to construct
solutions with a specific scaling dimension starting from an appropriate
solution of the bulk form factor axioms \cite{bffexp}. 

We remark that using the bulk form factor bootstrap (cf. \cite{Smirnov}
for a review) as a guide it is straightforward to extend these axioms
for non-diagonal scattering, i.e. particles with an internal degree
of freedom. Some results for such theories (albeit only for diagonal
boundary scattering) can be found in \cite{SGff,lashkevich}.

\section{A conjecture for the expectation values}

Consider a theory with a spectrum that contains a single massive particle
species of mass $m$. Leclair and Mussardo proposed the following
expression for the bulk finite temperature one-point functions \cite{leclairmussardo}:\begin{equation}
\langle\mathcal{A}\rangle^{R}=\sum_{n=0}^{\infty}\frac{1}{n!}\prod_{i=1}^{n}\left(\int_{-\infty}^{\infty}\frac{d\theta_{i}}{2\pi}\frac{\mathrm{e}^{-\epsilon(\theta_{i})}}{1+\mathrm{e}^{-\epsilon(\theta_{i})}}\right)f_{2n}^{c}(\theta_{1},...,\theta_{n})\label{eq:leclmuss1pt}\end{equation}
where $f_{2n}^{c}$ is the connected diagonal form factor of the local
bulk operator $\mathcal{A}$, $R=1/T$ in terms of the temperature
$T$, and $\epsilon(\theta)$ is the pseudo-energy function, which
is the solution of the thermodynamic Bethe Ansatz (TBA) equation

\begin{equation}
\epsilon(\theta)=mR\cosh\theta-\int_{-\infty}^{\infty}\frac{d\theta'}{2\pi}\varphi(\theta-\theta')\log(1+\mathrm{e}^{-\epsilon(\theta')})\label{eq:TBA}\end{equation}
where\[
\varphi(\theta)=\frac{d}{d\theta}\delta(\theta)\]
is the derivative of the two-particle phase-shift introduced in (\ref{eq:tpphasehsift}).
The factor $1/n!$ takes into account the fact that a complete set
of $n$-particle in-states is obtained with the ordering $\theta_{1}\geq\theta_{2}\geq\dots\geq\theta_{n}$,
but the integrals can be extended to the entire space using the fact
that the functions $f_{2n}^{c}(\theta_{1},...,\theta_{n})$ are symmetric
in all of their arguments.

The main idea behind the formula (\ref{eq:leclmuss1pt}) comes from
the TBA expression of the free energy \[
f(R)=-\int_{-\infty}^{\infty}\frac{d\theta}{2\pi}m\cosh(\theta)\log(1+\mathrm{e}^{-\epsilon(\theta)})\]
which shows that the finite temperature vacuum can be considered as
a free Fermi gas of quasi-particles for which the thermal weight is
given by the pseudo-energy function $\epsilon(\theta)$. The essential
condition necessary for the validity of this picture is that the complete
set of states used to derive (\ref{eq:leclmuss1pt}) must be inserted
at a position which is asymptotically far from any local operator
insertion, so that their distribution is governed by the unperturbed
finite temperature ground state. This is the reason why the Leclair-Mussardo
conjecture does not work for the two-point functions \cite{saleurfiniteT},
because the states inserted between the two local operators cannot
be asymptotically far from the positions of the operators which are
themselves located at a finite distance from each other.

From figure \ref{fig:infinitecylinder} it is obvious that a complete
set of asymptotic states can be inserted at $x=-\infty$ where their
distribution is unaffected by the presence of the boundary operator
$\mathcal{O}$. The only difference to the bulk case is that the complete
system of in-states is spanned by multi-particle states with all their
rapidities positive (i.e. with all particles moving towards the boundary),
so the natural generalization of (\ref{eq:leclmuss1pt}) is \begin{equation}
\langle\mathcal{O}\rangle^{R}=\sum_{n=0}^{\infty}\frac{1}{n!}\prod_{i=1}^{n}\left(\int_{0}^{\infty}\frac{d\theta_{i}}{2\pi}\frac{\mathrm{e}^{-\epsilon(\theta_{i})}}{1+\mathrm{e}^{-\epsilon(\theta_{i})}}\right)F_{2n}^{c}(\theta_{1},...,\theta_{n})\label{eq:myconjecture}\end{equation}
where $F_{2n}^{c}$ is the connected part of the diagonal form factor
of the local boundary operator $\mathcal{O}$:\begin{equation}
F_{2n}^{c}(\theta_{1},...,\theta_{n})=\,\langle\theta_{1},\theta_{2},\dots,\theta_{n}\vert\mathcal{O}(t=0)\vert\theta_{1},\theta_{2},\dots,\theta_{n}\rangle^{connected}\label{eq:connectedsimplenotation}\end{equation}
which is again symmetric in all their variables as a result of equation
(\ref{eq:permutation}). The precise definition of the connected matrix
element (valid both for bulk and the boundary operators) is specified
later in subsection 4.2. 

The conjectured expression (\ref{eq:myconjecture}) can be checked
against a calculation of the low-temperature expansion using the boundary
form factors; this calculation is performed in the sequel. However,
the kinematical residue equation (\ref{eq:kinematical}) implies that
diagonal matrix elements contain disconnected terms which are infinite,
and therefore must be regularized. As shown in \cite{fftcsa2} a natural
regularization can be obtained by putting the system in a finite volume,
which was implemented for the bulk case in \cite{fftcsa1,fftcsa2}
and for the boundary case in \cite{bfftcsa}.

For completeness we note that the conjecture (\ref{eq:myconjecture})
can be extended to a theory with multiple particle species and diagonal
scattering in the following form:\begin{equation}
\langle\mathcal{O}\rangle^{R}=\sum_{n=0}^{\infty}\frac{1}{n!}\sum_{a_{1}}\dots\sum_{a_{n}}\prod_{i=1}^{n}\left(\int_{0}^{\infty}\frac{d\theta_{i}}{2\pi}\frac{\mathrm{e}^{-\epsilon_{a_{i}}(\theta_{i})}}{1+\mathrm{e}^{-\epsilon_{a_{i}}(\theta_{i})}}\right)F_{a_{1}\dots a_{n}}^{c}(\theta_{1},...,\theta_{n})\label{eq:genconjecture}\end{equation}
where \[
F_{a_{1}\dots a_{n}}^{c}(\theta_{1},...,\theta_{n})=\,_{a_{1}\dots a_{n}}\langle\theta_{1},\theta_{2},\dots,\theta_{n}\vert\mathcal{O}(t=0)\vert\theta_{1},\theta_{2},\dots,\theta_{n}\rangle_{a_{1}\dots a_{n}}^{connected}\]
while the pseudo-energy functions satisfy \[
\epsilon_{a}(\theta)=m_{a}R\cosh\theta-\sum_{b}\int\frac{d\theta'}{2\pi}\varphi_{ab}(\theta-\theta')\log(1+\mathrm{e}^{-\epsilon_{ab}(\theta')})\]
where\begin{equation}
\varphi_{ab}(\theta)=\frac{d}{d\theta}\delta_{ab}(\theta)\label{eq:derbulkphaseshift}\end{equation}
are the derivatives of the two-particle phase-shifts introduced in
(\ref{eq:tpphasehsift}). For the sake of simplicity the species labels
will be omitted from now on, i.e. every formula will be written for
the case of a single particle species; the extension to multiple species
(with diagonal scattering) is rather straightforward. 

Eqn. (\ref{eq:myconjecture}) can be expanded systematically order
by order in $\mathrm{e}^{-mR}$ which yields a low temperature expansion,
following the procedure implemented for the Leclair-Mussardo formula
(\ref{eq:leclmuss1pt}) in \cite{fftcsa2}. The detailed calculation
is performed in Appendix A with the following result: \begin{equation}
\langle\mathcal{O}\rangle^{R}=\sigma_{1}+\sigma_{2}+\sigma_{3}+O\left(\mbox{e}^{-4mR}\right)\label{eq:myconjexpanded}\end{equation}
where\begin{eqnarray*}
\sigma_{1} & = & \int_{0}^{\infty}\frac{d\theta_{1}}{2\pi}\left(\mathrm{e}^{-mR\cosh\theta_{1}}-\mathrm{e}^{-2mR\cosh\theta_{1}}+\mathrm{e}^{-3mR\cosh\theta_{1}}\right)F_{2}^{c}(\theta_{1})\\
 &  & +\frac{1}{2}\int_{0}^{\infty}\frac{d\theta_{1}}{2\pi}\int_{0}^{\infty}\frac{d\theta_{2}}{2\pi}\mathrm{e}^{-mR(\cosh\theta_{1}+\cosh\theta_{2})}\Phi_{12}\left(F_{2}^{c}(\theta_{1})+F_{2}^{c}(\theta_{2})\right)\\
 &  & +\frac{1}{2}\int_{0}^{\infty}\frac{d\theta_{1}}{2\pi}\int_{0}^{\infty}\frac{d\theta_{2}}{2\pi}\int_{0}^{\infty}\frac{d\theta_{3}}{2\pi}\mathrm{e}^{-mR(\cosh\theta_{1}+\cosh\theta_{2}+\cosh\theta_{3})}\big(\Phi_{12}\Phi_{13}\\
 &  & +\Phi_{12}\Phi_{23}+\Phi_{13}\Phi_{23}\big)F_{2}^{c}(\theta_{3})\\
 &  & -\int_{0}^{\infty}\frac{d\theta_{1}}{2\pi}\int_{0}^{\infty}\frac{d\theta_{2}}{2\pi}\mathrm{e}^{-mR(2\cosh\theta_{1}+\cosh\theta_{2})}\left(2F_{2}^{c}(\theta_{1})+\frac{1}{2}F_{2}^{c}(\theta_{2})\right)\Phi_{12}\\
\sigma_{2} & = & \frac{1}{2}\int_{0}^{\infty}\frac{d\theta_{1}}{2\pi}\int_{0}^{\infty}\frac{d\theta_{2}}{2\pi}\mathrm{e}^{-mR(\cosh\theta_{1}+\cosh\theta_{2})}F_{4}^{c}(\theta_{1},\theta_{2})\\
 &  & -\int_{0}^{\infty}\frac{d\theta_{1}}{2\pi}\int_{0}^{\infty}\frac{d\theta_{2}}{2\pi}\mathrm{e}^{-mR(2\cosh\theta_{1}+\cosh\theta_{2})}F_{4}^{c}(\theta_{1},\theta_{2})\\
 &  & +\frac{1}{2}\int_{0}^{\infty}\frac{d\theta_{1}}{2\pi}\int_{0}^{\infty}\frac{d\theta_{2}}{2\pi}\int_{0}^{\infty}\frac{d\theta_{3}}{2\pi}\mathrm{e}^{-mR(\cosh\theta_{1}+\cosh\theta_{2}+\cosh\theta_{3})}\left(\Phi_{12}+\Phi_{13}\right)F_{4}^{c}(\theta_{2},\theta_{3})\\
\sigma_{3} & = & \frac{1}{6}\int_{0}^{\infty}\frac{d\theta_{1}}{2\pi}\int_{0}^{\infty}\frac{d\theta_{2}}{2\pi}\int_{0}^{\infty}\frac{d\theta_{3}}{2\pi}\mathrm{e}^{-mR(\cosh\theta_{1}+\cosh\theta_{2}+\cosh\theta_{3})}F_{6}^{c}(\theta_{1},\theta_{2},\theta_{3})\end{eqnarray*}
and \[
\Phi_{ij}=\varphi(\theta_{i}-\theta_{j})+\varphi(\theta_{i}+\theta_{j})\]
For later convenience some terms were reordered by reshuffling the
integral variables. 

In the sequel this result is compared to the result obtained from
explicit evaluation of the finite temperature Gibbs average. In order
to perform this calculation it is necessary to use finite volume as
a regulator, and so now we turn to the issue of boundary form factors
in finite volume, based on the results of \cite{bfftcsa}.

\section{Boundary form factors in finite volume}

\subsection{Bethe-Yang equations}

\begin{figure}
\begin{centering}
\psfrag{xinf}{$x$}
\psfrag{R}{$R=1/T$}
\psfrag{O}{$\mathcal{O}$}
\psfrag{x=L}{$x=-L$}
\psfrag{a}{$\alpha$}
\psfrag{b}{$\beta$}
\psfrag{x=0}{$x=0$}\includegraphics{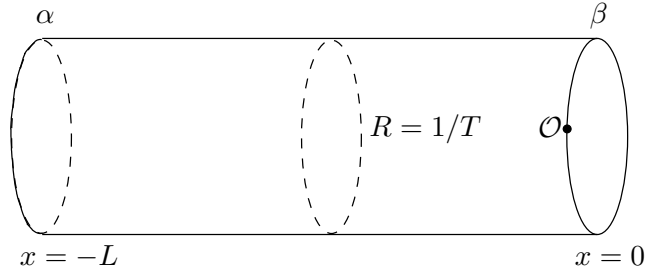}
\par\end{centering}

\caption{\label{fig:finvolcylinder} The setting of Fig. \ref{fig:infinitecylinder}
in finite volume}

\end{figure}

Let us consider an integrable boundary quantum field theory with particles
of species $a=1,\dots,N$ and corresponding masses $m_{a}$ in finite
volume $L$ as shown in figure \ref{fig:finvolcylinder}. As in section
2, the bulk and boundary scattering is assumed to be diagonal and
given by the two-particle $S$ matrices\[
S_{a_{1}a_{2}}\left(\theta_{1}-\theta_{2}\right)=\mbox{e}^{i\delta_{a_{1}a_{2}}\left(\theta_{1}-\theta_{2}\right)}\]
and the one-particle reflection factors \begin{equation}
R_{a}^{(\alpha)}\left(\theta\right)=\mbox{e}^{i\delta_{a}^{(\alpha)}\left(\theta\right)}\qquad,\qquad R_{a}^{(\beta)}\left(\theta\right)=\mbox{e}^{i\delta_{a}^{(\beta)}\left(\theta\right)}\label{eq:bphaseshifts}\end{equation}
where $\alpha$ and $\beta$ denote the left and right boundary conditions,
respectively.

In the diagonal case, the multi-particle energy levels in a finite
volume $L$ are described by the following Bethe-Yang equations \cite{bbye}:\begin{eqnarray}
Q_{j}\left(\theta_{1},\dots,\theta_{n}\right)_{a_{1}\dots a_{n}} & = & 2\pi I_{j}\label{eq:bbye}\end{eqnarray}
where the phases describing the wave function monodromies are\begin{eqnarray*}
Q_{j}\left(\theta_{1},\dots,\theta_{n}\right)_{a_{1}\dots a_{n}} & = & 2m_{a_{j}}L\sinh\theta_{j}+\delta_{a_{j}}^{(\alpha)}\left(\theta_{j}\right)+\delta_{a_{j}}^{(\beta)}\left(\theta_{j}\right)\\
 &  & +\sum_{k\neq j}\left(\delta_{a_{j}a_{k}}\left(\theta_{j}-\theta_{k}\right)+\delta_{a_{j}a_{k}}\left(\theta_{j}+\theta_{k}\right)\right)\end{eqnarray*}
Here all rapidities $\theta_{j}$ (and accordingly all quantum numbers
$I_{j}$) are taken to be positive%
\footnote{Boundary reflections change the sign of the momentum, so finite volume
multi-particle states can be characterized by the absolute value of
the rapidities.%
}. The corresponding multi-particle state is denoted by \[
\vert\{I_{1},\dots,I_{n}\}\rangle_{a_{1}\dots a_{n},L}\]
and its energy (relative to the ground state) is\[
E_{I_{1}\dots I_{n}}(L)=\sum_{j=1}^{n}m_{a_{j}}\cosh\tilde{\theta}_{j}\]
where $\left\{ \tilde{\theta}_{j}\right\} _{j=1,\dots,n}$ is the
solution of eqns. (\ref{eq:bbye}) in volume $L$. The energy calculated
from the Bethe-Yang equations is exact to all order in $1/L$; only
finite size effects decaying exponentially with $L$ are neglected.

\subsection{Matrix elements in finite volume}

In general infinite volume and finite volume matrix elements are just
related by the square root of the ratio of normalization of the corresponding
states \cite{bfftcsa,fftcsa1}. This results in the following relation:\begin{eqnarray}
 &  & \,_{b_{1}\dots b_{m}}\langle\{I_{1}',\dots,I_{m}'\}\vert\mathcal{O}(0)\vert\{I_{1},\dots,I_{n}\}\rangle_{a_{1}\dots a_{n},L}=\nonumber \\
 &  & \qquad\frac{F_{\bar{b}_{m}\dots\bar{b}_{1}a_{1}\dots a_{n}}^{\mathcal{O}}(\tilde{\theta}_{m}'+i\pi,\dots,\tilde{\theta}_{1}'+i\pi,\tilde{\theta}_{1},\dots,\tilde{\theta}_{n})}{\sqrt{\rho_{a_{1}\dots a_{n}}(\tilde{\theta}_{1},\dots,\tilde{\theta}_{n})\rho_{b_{1}\dots b_{m}}(\tilde{\theta}_{1}',\dots,\tilde{\theta}_{m}')}}+O(\mathrm{e}^{-\mu L})\label{eq:genffrelation}\end{eqnarray}
where $F_{a_{1}\dots a_{n}}^{\mathcal{O}}(\tilde{\theta}_{1},\dots,\tilde{\theta}_{n})$
is the form factor of the operator $\mathcal{O}$ (in the infinite
volume theory, i.e. on the half-line $x<0$),$\left\{ \tilde{\theta}_{j}\right\} _{j=1,\dots,n}$
is the solution of eqns. (\ref{eq:bbye}) in volume $L$ for the set
of quantum numbers $\{I_{1},\dots,I_{n}\}$ (similarly for $\left\{ \tilde{\theta}_{j}'\right\} _{j=1,\dots,m}$
and $\{I_{1}',\dots,I_{m}'\}$), and \begin{equation}
\rho_{a_{1}\dots a_{n}}(\theta_{1},\dots,\theta_{n})=\det\left\{ \frac{\partial Q_{k}(\theta_{1},\dots,\theta_{n})_{a_{1}\dots a_{n}}}{\partial\theta_{l}}\right\} _{k,l=1,\dots,n}\label{eq:bydet}\end{equation}
is the finite volume density of states, which is the Jacobi determinant
of the mapping between the space of quantum numbers and the space
of rapidities specified by the Bethe-Yang equations (\ref{eq:bbye}).
An explicit expression for the derivative matrix of the Bethe-Yang
equations (\ref{eq:bbye}) is \begin{eqnarray}
\frac{\partial Q_{k}}{\partial\theta_{k}} & = & 2m_{a_{k}}L\cosh\theta_{k}+\psi_{a_{k}}^{(\alpha)}(\theta_{k})+\psi_{a_{k}}^{(\beta)}(\theta_{k})+\sum_{j\neq k}[\varphi_{a_{j}a_{k}}(\theta_{j}-\theta_{k})+\varphi_{a_{j}a_{k}}(\theta_{j}+\theta_{k})]\nonumber \\
\frac{\partial Q_{k}}{\partial\theta_{j}} & = & -\varphi_{a_{j}a_{k}}(\theta_{j}-\theta_{k})+\varphi_{a_{j}a_{k}}(\theta_{j}+\theta_{k})\qquad,\qquad j\neq k\label{eq:jacmatexplicit}\end{eqnarray}
where \[
\psi_{a}^{(\alpha)}(\theta)=\frac{d}{d\theta}\delta_{a}^{(\alpha)}(\theta)\;,\;\psi^{(\beta)}(\theta)=\frac{d}{d\theta}\delta_{a}^{(\beta)}(\theta)\]
are the derivatives of the boundary phase-shifts defined in (\ref{eq:bphaseshifts}),
while the $\varphi$ are the derivatives of the bulk ones as written
in (\ref{eq:derbulkphaseshift}). 

Eqn. (\ref{eq:genffrelation}) is valid as long as the sets of the
rapidities corresponding to the two states, $\left\{ \tilde{\theta}_{j}\right\} _{j=1,\dots,n}$
and $\left\{ \tilde{\theta}_{j}'\right\} _{j=1,\dots,m}$, are disjoint
i.e. when there are no disconnected contributions. For diagonal matrix
elements\[
\,_{a_{1}\dots a_{n}}\langle\{I_{1},\dots,I_{n}\}\vert\mathcal{O}(0)\vert\{I_{1},\dots,I_{n}\}\rangle_{a_{1}\dots a_{n},L}\]
 a more careful analysis is required \cite{fftcsa2,bfftcsa}. According
to (\ref{eq:genffrelation}) for this case it is necessary to consider
\[
F_{\bar{a}_{n}\dots\bar{a}_{1}a_{1}\dots a_{n}}(\theta_{n}+i\pi,...,\theta_{1}+i\pi,\theta_{1},...,\theta_{n})\]
Because of the kinematical poles the above expression is not well-defined.
The bulk kinematical singularity axiom (\ref{eq:kinematical}) implies
that the regularized version\[
F_{\bar{a}_{n}\dots\bar{a}_{1}a_{1}\dots a_{n}}(\theta_{n}+i\pi+\epsilon_{n},...,\theta_{1}+i\pi+\epsilon_{1},\theta_{1},...,\theta_{n})\]
has a finite limit when $\epsilon_{i}\rightarrow0$ simultaneously.
However, the end result depends on the direction of the limit, i.e.
on the ratio of the $\epsilon_{i}$ parameters. The terms that are
relevant in this limit can be written in the following general form:
\begin{eqnarray}
F_{\bar{a}_{n}\dots\bar{a}_{1}a_{1}\dots a_{n}}(\theta_{n}+i\pi+\epsilon_{n},...,\theta_{1}+i\pi+\epsilon_{1},\theta_{1},...,\theta_{n})=\label{mostgeneral}\\
\prod_{i=1}^{n}\frac{1}{\epsilon_{i}}\cdot\sum_{i_{1}=1}^{n}...\sum_{i_{n}=1}^{n}\mathcal{A}_{i_{1}...i_{n}}^{a_{1}\dots a_{n}}(\theta_{1},\dots,\theta_{n})\epsilon_{i_{1}}\epsilon_{i_{2}}...\epsilon_{i_{n}}+\dots\nonumber \end{eqnarray}
 where $\mathcal{A}_{i_{1}...i_{n}}^{a_{1}\dots a_{n}}$ is a tensor
of rank $n$ in the indices $i_{1},\dots,i_{n}$ which is symmetric
under the exchange of indices that correspond to particles of the
same species, and the ellipsis denote terms that vanish when taking
$\epsilon_{i}\rightarrow0$ simultaneously. 

The connected matrix element can be defined as the $\epsilon_{i}$
independent part of eqn. (\ref{mostgeneral}), i.e. the part which
does not diverge whenever any of the $\epsilon_{i}$ is taken to zero:
\begin{equation}
F_{a_{1}\dots a_{n}}^{c}(\theta_{1},...,\theta_{n})=\sum_{(p_{1}\dots p_{n})}\mathcal{A}_{p_{1}\dots p_{n}}^{a_{1}\dots a_{n}}(\theta_{1},\dots,\theta_{n})\label{eq:connected}\end{equation}
where the summation goes over all permutations $(p_{1},\dots,p_{n})$
of the numbers $1,\dots,n$. As shown in appendix B, all other evaluations
of the diagonal matrix elements (\ref{mostgeneral}) can be readily
expressed in terms of the connected amplitudes. 

It was shown in \cite{bfftcsa} that a natural generalization of an
expression proposed earlier by Saleur \cite{saleurfiniteT} for bulk
diagonal matrix elements can be extended to the boundary case in the
following way%
\footnote{Note that here the original result of \cite{bfftcsa} is extended
to the case of several particle species.%
}:\begin{eqnarray}
 &  & \,_{a_{1}\dots a_{n}}\langle\{I_{1}\dots I_{n}\}|\mathcal{O}(0)|\{I_{1}\dots I_{n}\}\rangle_{a_{1}\dots a_{n},L}=\label{eq:diaggenrulesaleur}\\
 &  & \frac{1}{\rho_{a_{1}\dots a_{n}}(\tilde{\theta}_{1},\dots,\tilde{\theta}_{n})}\sum_{A\subset\{1,2,\dots n\}}F_{a(A)}^{c}(\{\tilde{\theta}_{k}\}_{k\in A})\tilde{\rho}_{a_{1}\dots a_{n}}(\tilde{\theta}_{1},\dots,\tilde{\theta}_{n}|A)+O(\mathrm{e}^{-\mu L})\nonumber \end{eqnarray}
The summation runs over all subsets $A$ of $\{1,2,\dots n\}$ and
again $\left\{ \tilde{\theta}_{j}\right\} _{j=1,\dots,n}$ is the
solution of eqns. (\ref{eq:bbye}) in volume $L$ for the set of quantum
numbers $\{I_{1},\dots,I_{n}\}$. For any such subset the corresponding
species index list is defined as \[
a(A)=\left\{ a_{k}\right\} _{k\in A}\]
and \begin{equation}
\tilde{\rho}_{a_{1}\dots a_{n}}(\theta_{1},\dots,\theta_{n}|A)=\det\mathcal{J}_{A}^{a_{1}\dots a_{n}}(\theta_{1},\dots,\theta_{n})\label{eq:rhotilde_def}\end{equation}
is the appropriate sub-determinant of the $n\times n$ Bethe-Yang
Jacobi matrix \begin{equation}
\mathcal{J}_{a_{1}\dots a_{n}}(\theta_{1},\dots,\theta_{n})_{kl}=\frac{\partial Q_{k}(\theta_{1},\dots,\theta_{n})_{a_{1}\dots a_{n}}}{\partial\theta_{l}}\label{eq:jacsubmat}\end{equation}
obtained by deleting the rows and columns corresponding to the subset
of indices $A$. The determinant of the empty sub-matrix (i.e. when
$A=\{1,2,\dots n\}$) is defined to equal $1$ by convention. It is
also shown in appendix B that the symmetric evaluation which gave
a very convenient alternative to (\ref{eq:diaggenrulesaleur}) in
the bulk \cite{fftcsa2}, behaves rather differently in the boundary
case.

\section{Expansion of finite temperature expectation values}

\subsection{Low-temperature expansion for one-point functions}

The procedure leading to a well-defined low-temperature expansion
was outlined in section 7 of \cite{fftcsa2}; details about the validity
of the method and the existence of the limits taken are omitted (the
interested reader is referred to the above paper for details). Let
us evaluate the finite temperature expectation value of an operator
$\mathcal{O}$ located at $x=0$ in a finite but large volume $L$,
according to the setting introduced in section 4:

\begin{equation}
\langle\mathcal{O}\rangle_{L}^{R}=\frac{\text{Tr}_{L}\left(\mathrm{e}^{-RH_{L}}\mathcal{O}\right)}{\text{Tr}_{L}\left(\mathrm{e}^{-RH_{L}}\right)}\qquad,\; T=1/R\label{onepointRL}\end{equation}
$H_{L}$ is the finite volume Hamiltonian, and $\mathrm{Tr}_{L}$
means that the trace is now taken over the finite volume Hilbert space.
The expectation value $\left\langle \mathcal{O}\right\rangle ^{R}$
can be recovered in the limit $L\,\rightarrow\,\infty$ which means
that the left boundary condition $\alpha$ in figure \ref{fig:finvolcylinder}
plays an auxiliary role, and the end result can only depend on the
$x=0$ boundary condition $\beta$; this issue will be taken up again
in subsection 5.5.

In the calculation below particle species labels are dropped for simplicity
(they can be easily reinstated if necessary) and we use the simplified
notation $F_{2n}$ for the $n$-particle diagonal matrix element introduced
in (\ref{eq:connectedsimplenotation}). It is also convenient to introduce
a new notation:\[
|\theta_{1},\dots,\theta_{n}\rangle_{L}=|\{I_{1},\dots,I_{n}\}\rangle_{L}\]
where $\theta_{1},\dots,\theta_{n}$ solve the Bethe-Yang equations
(\ref{eq:bbye}) for $n$ particles with quantum numbers $I_{1},\dots,I_{n}$
in volume $L$; as remarked in subsection 2.1, all of the rapidities
can be taken positive. The low temperature expansion of (\ref{onepointRL})
can be developed in orders of $\mathrm{e}^{-mR}$ using

\begin{eqnarray}
\text{Tr}_{L}\left(\mathrm{e}^{-RH_{L}}\mathcal{O}\right) & = & \langle\mathcal{O}\rangle_{L}+\sum_{\theta^{(1)}}\mathrm{e}^{-mR\cosh\theta^{(1)}}\langle\theta^{(1)}|\mathcal{O}|\theta^{(1)}\rangle_{L}\nonumber \\
 &  & +\frac{1}{2}\sum_{\theta_{1}^{(2)},\theta_{2}^{(2)}}{}^{'}\mathrm{e}^{-mR(\cosh\theta_{1}^{(2)}+\cosh\theta_{2}^{(2)})}\langle\theta_{1}^{(2)},\theta_{2}^{(2)}|\mathcal{O}|\theta_{1}^{(2)},\theta_{2}^{(2)}\rangle_{L}+\nonumber \\
 &  & +\frac{1}{6}\sum_{\theta_{1}^{(3)},\theta_{2}^{(3)},\theta_{3}^{(3)}}{}^{'}\mathrm{e}^{-mR(\cosh\theta_{1}^{(3)}+\cosh\theta_{2}^{(3)}+\cosh\theta_{3}^{(3)})}\langle\theta_{1}^{(3)},\theta_{2}^{(3)},\theta_{3}^{(3)}|\mathcal{O}|\theta_{1}^{(3)},\theta_{2}^{(3)},\theta_{3}^{(3)}\rangle_{L}\nonumber \\
 &  & +O(\mathrm{e}^{-4mR})\label{eq:nomexp}\end{eqnarray}
and\begin{eqnarray}
\text{Tr}_{L}\left(\mathrm{e}^{-RH_{L}}\right) & = & 1+\sum_{\theta^{(1)}}\mathrm{e}^{-mR\cosh(\theta^{(1)})}+\frac{1}{2}\sum_{\theta_{1}^{(2)},\theta_{2}^{(2)}}{}^{'}\mathrm{e}^{-mR(\cosh(\theta_{1}^{(2)})+\cosh(\theta_{2}^{(2)}))}\nonumber \\
 &  & +\frac{1}{6}\sum_{\theta_{1}^{(3)},\theta_{2}^{(3)},\theta_{3}^{(3)}}{}^{'}\mathrm{e}^{-mR(\cosh\theta_{1}^{(3)}+\cosh\theta_{2}^{(3)}+\cosh\theta_{3}^{(3)})}+O(\mathrm{e}^{-4mR})\label{eq:Zexp}\end{eqnarray}
The denominator of (\ref{onepointRL}) can then be easily expanded:\begin{eqnarray}
\frac{1}{\text{Tr}_{L}\left(\mathrm{e}^{-RH_{L}}\right)} & = & 1-\sum_{\theta^{(1)}}\mathrm{e}^{-mR\cosh\theta^{(1)}}+\left(\sum_{\theta^{(1)}}\mathrm{e}^{-mR\cosh\theta^{(1)}}\right)^{2}-\frac{1}{2}\sum_{\theta_{1}^{(2)},\theta_{2}^{(2)}}{}^{'}\mathrm{e}^{-mR(\cosh\theta_{1}^{(2)}+\cosh\theta_{2}^{(2)})}\nonumber \\
 &  & -\left(\sum_{\theta^{(1)}}\mathrm{e}^{-mR\cosh\theta^{(1)}}\right)^{3}+\left(\sum_{\theta^{(1)}}\mathrm{e}^{-mR\cosh\theta^{(1)}}\right)\sum_{\theta_{1}^{(2)},\theta_{2}^{(2)}}{}^{'}\mathrm{e}^{-mR(\cosh\theta_{1}^{(2)}+\cosh\theta_{2}^{(2)})}\nonumber \\
 &  & -\frac{1}{6}\sum_{\theta_{1}^{(3)},\theta_{2}^{(3)},\theta_{3}^{(3)}}{}^{'}\mathrm{e}^{-mR(\cosh\theta_{1}^{(3)}+\cosh\theta_{2}^{(3)}+\cosh\theta_{3}^{(3)})}+O(\mathrm{e}^{-4mR})\label{eq:Zinvexp}\end{eqnarray}
The primes in the multi-particle sums serve as a reminder that there
exist only states for which all quantum numbers are distinct. Since
it was assumed that there is a single particle species, this means
that terms in which any two of the rapidities coincide are excluded.
All $n$-particle terms in (\ref{eq:nomexp}) and (\ref{eq:Zexp})
have a $1/n!$ prefactor which takes into account that different ordering
of the same rapidities give the same state; as the expansion contains
only diagonal matrix elements, phases resulting from reordering the
particles cancel. It is also crucial to remember that in the boundary
case the summations only run over positive values of the rapidities
(cf. section 3). The upper indices of the rapidity variables indicate
the number of particles in the original finite volume states which
helps to keep track which multi-particle state density is relevant. 

It is also necessary to extend the finite volume matrix elements to
rapidities that are not necessarily solutions of the appropriate Bethe-Yang
equations. The required analytic continuation can be written down
using eqn. (\ref{eq:diaggenrulesaleur}):

\begin{equation}
\langle\theta_{1},\dots,\theta_{n}|\mathcal{O}|\theta_{1},\dots,\theta_{n}\rangle_{L}=\frac{1}{\rho_{n}(\theta_{1},\dots,\theta_{n})_{L}}\,\sum_{A\subset\{1,2,\dots n\}}F_{2|A|}^{c}(\{\theta_{i}\}_{i\in A})\tilde{\rho}(\theta_{1},\dots,\theta_{n}|A)_{L}+O(\mathrm{e}^{-\mu L})\label{eq:Fcextended}\end{equation}
where the volume dependence of the $n$-particle density factors was
made explicit and the form factors are computed from solutions of
the bootstrap equations in section 2 with the boundary condition $\beta$.
It is apparent that the continuation is specified only up to terms
decaying exponentially with the volume $L$ but this is sufficient
for the evaluation of the $L\,\rightarrow\,\infty$ limit of (\ref{onepointRL}).

It is useful to notice that unitarity and real analyticity imply that
all the phase-shift derivatives\begin{equation}
\varphi(\theta)=\frac{d}{d\theta}\delta(\theta)\;,\;\psi^{(\alpha)}(\theta)=\frac{d}{d\theta}\delta^{(\alpha)}(\theta)\;,\;\psi^{(\beta)}(\theta)=\frac{d}{d\theta}\delta^{(\beta)}(\theta)\label{eq:dreflphaseshift}\end{equation}
are real and even functions. Another important observation is that
the exclusion principle (cf. section 2) implies that the amplitudes
$F_{2n}^{c}(\theta_{1},\dots,\theta_{n})$ vanish whenever any two
of their rapidity arguments coincide. In addition, the connected form
factor functions are symmetric under permutations of their arguments
according to their definition (\ref{eq:connected}) and (in contrast
to the bulk case) they are even functions in all of their rapidity
arguments separately i.e.\[
F_{2n}^{c}(\theta_{1},\theta_{2},\dots,\theta_{n})=F_{2n}^{c}(-\theta_{1},\theta_{2},\dots,\theta_{n})\]
which is a result of the reflection equation (\ref{eq:reflection})
satisfied by the form factors.

\subsection{Lowest order terms}

The leading correction is \[
\langle\mathcal{O}\rangle_{L}^{R}=\langle\mathcal{O}\rangle_{L}+\sum_{\theta^{(1)}}\mathrm{e}^{-mR\cosh\theta^{(1)}}\left(\langle\theta^{(1)}|\mathcal{O}|\theta^{(1)}\rangle_{L}-\langle\mathcal{O}\rangle_{L}\right)+O(\mathrm{e}^{-2mR})\]
From (\ref{eq:Fcextended})\[
\langle\theta|\mathcal{O}|\theta\rangle_{L}-\langle\mathcal{O}\rangle=\frac{1}{\rho_{1}(\theta)}F_{2}^{c}(\theta)+O\left(\mathrm{e}^{-\mu L}\right)\]
Note also that the difference between the finite volume vacuum expectation
value and the infinite volume one decays exponentially with $L$ \[
\langle\mathcal{O}\rangle_{L}-\langle\mathcal{O}\rangle\sim O\left(\mathrm{e}^{-\mu L}\right)\]
From now on such exponential corrections will simply be omitted. In
the large $L$ limit the summation can be replaced by the integral\[
\sum_{\theta^{(1)}}\rightarrow\int\frac{d\theta}{2\pi}\rho_{1}(\theta)\]
and therefore \begin{equation}
\langle\mathcal{O}\rangle^{R}=\langle\mathcal{O}\rangle+\int_{0}^{\infty}\frac{d\theta}{2\pi}F_{2}^{c}(\theta)\mathrm{e}^{-mR\cosh\theta}+O(\mathrm{e}^{-2mR})\label{eq:bff1order}\end{equation}

\subsection{Corrections of order $\mathrm{e}^{-2mR}$}

To this order one has 

\begin{eqnarray*}
\langle\mathcal{O}\rangle_{L}^{R} & = & \langle\mathcal{O}\rangle_{L}+\sum_{\theta^{(1)}}\mathrm{e}^{-mR\cosh\theta^{(1)}}\left(\langle\theta^{(1)}|\mathcal{O}|\theta^{(1)}\rangle_{L}-\langle\mathcal{O}\rangle_{L}\right)\\
 &  & -\left(\sum_{\theta_{1}^{(1)}}\mathrm{e}^{-mR\cosh\theta_{1}^{(1)}}\right)\left(\sum_{\theta_{2}^{(1)}}\mathrm{e}^{-mR\cosh\theta_{2}^{(1)}}\left(\langle\theta_{2}^{(1)}|\mathcal{O}|\theta_{2}^{(1)}\rangle_{L}-\langle\mathcal{O}\rangle_{L}\right)\right)\\
 &  & +\frac{1}{2}\sum_{\theta_{1}^{(2)},\theta_{2}^{(2)}}{}^{'}\mathrm{e}^{-mR(\cosh\theta_{1}^{(2)}+\cosh\theta_{2}^{(2)})}\left(\langle\theta_{1}^{(2)},\theta_{2}^{(2)}|\mathcal{O}|\theta_{1}^{(2)},\theta_{2}^{(2)}\rangle_{L}-\langle\mathcal{O}\rangle_{L}\right)+O(\mathrm{e}^{-3mR})\end{eqnarray*}
Using the symmetry of the first term in the rapidities and separating
the diagonal contribution from the double summation on the last line
leads to

\begin{eqnarray*}
 &  & -\frac{1}{2}\left(\sum_{\theta_{1}^{(1)}}\mathrm{e}^{-mR\cosh\theta_{1}^{(1)}}\right)\left(\sum_{\theta_{2}^{(1)}}\mathrm{e}^{-mR\cosh\theta_{2}^{(1)}}\left(\langle\theta_{1}^{(1)}|\mathcal{O}|\theta_{1}^{(1)}\rangle_{L}+\langle\theta_{2}^{(1)}|\mathcal{O}|\theta_{2}^{(1)}\rangle_{L}-2\langle\mathcal{O}\rangle_{L}\right)\right)\\
 &  & +\frac{1}{2}\sum_{\theta_{1}^{(2)},\theta_{2}^{(2)}}\mathrm{e}^{-mR(\cosh\theta_{1}^{(2)}+\cosh\theta_{2}^{(2)})}\left(\langle\theta_{1}^{(2)},\theta_{2}^{(2)}|\mathcal{O}|\theta_{1}^{(2)},\theta_{2}^{(2)}\rangle_{L}-\langle\mathcal{O}\rangle_{L}\right)\\
 &  & -\frac{1}{2}\sum_{\theta_{1}^{(2)}=\theta_{2}^{(2)}}\mathrm{e}^{-2mR\cosh\theta_{1}^{(2)}}\left(\langle\theta_{1}^{(2)},\theta_{1}^{(2)}|\mathcal{O}|\theta_{1}^{(2)},\theta_{1}^{(2)}\rangle_{L}-\langle\mathcal{O}\rangle_{L}\right)\end{eqnarray*}
The terms containing two independent rapidity sums can be written
as\begin{eqnarray*}
\Sigma_{2}^{(2)} & = & \frac{1}{2}\int\frac{d\theta_{1}}{2\pi}\frac{d\theta_{2}}{2\pi}\mathrm{e}^{-mR(\cosh\theta_{1}+\cosh\theta_{2})}\Bigg(-\rho_{1}(\theta_{1})\rho_{1}(\theta_{2})\left(\frac{1}{\rho_{1}(\theta_{1})}F_{2}^{c}(\theta_{1})+\frac{1}{\rho_{1}(\theta_{2})}F_{2}^{c}(\theta_{2})\right)\\
 &  & +F_{4}^{c}(\theta_{1},\theta_{2})+\tilde{\rho}(\theta_{1},\theta_{2}|\{1\})F_{2}^{c}(\theta_{1})+\tilde{\rho}(\theta_{1},\theta_{2}|\{2\})F_{2}^{c}(\theta_{2})\Bigg)\end{eqnarray*}
where \[
\rho_{1}(\theta)=2mL\cosh\theta+\psi^{(\alpha)}(\theta)+\psi^{(\beta)}(\theta)\]
is the one-particle state density, while\begin{eqnarray}
\tilde{\rho}(\theta_{1},\theta_{2}|\{2\}) & = & 2mL\cosh\theta_{1}+\psi^{(\alpha)}(\theta_{1})+\psi^{(\beta)}(\theta_{1})+\varphi(\theta_{1}-\theta_{2})+\varphi(\theta_{1}+\theta_{2})\nonumber \\
\tilde{\rho}(\theta_{1},\theta_{2}|\{1\}) & = & 2mL\cosh\theta_{2}+\psi^{(\alpha)}(\theta_{2})+\psi^{(\beta)}(\theta_{2})+\varphi(\theta_{2}-\theta_{1})+\varphi(\theta_{2}+\theta_{1})\label{eq:rhotilde2_1}\end{eqnarray}
are the corresponding sub-determinants of the two-particle Bethe-Yang
Jacobian, evaluated according to (\ref{eq:rhotilde_def}). Taking
$L\,\rightarrow\,\infty$\[
\Sigma_{2}^{(2)}=\frac{1}{2}\int\frac{d\theta_{1}}{2\pi}\frac{d\theta_{2}}{2\pi}\mathrm{e}^{-mR(\cosh\theta_{1}+\cosh\theta_{2})}\left[F_{4}^{c}(\theta_{1},\theta_{2})+\left(\varphi(\theta_{1}-\theta_{2})+\varphi(\theta_{1}+\theta_{2})\right)\left(F_{2}^{c}(\theta_{1})+F_{2}^{c}(\theta_{2})\right)\right]\]
The diagonal contribution contains a single rapidity sum\[
\Sigma_{2}^{(1)}=-\frac{1}{2}\sum_{\theta_{1}^{(2)}=\theta_{2}^{(2)}}\mathrm{e}^{-2mR\cosh\theta_{1}^{(2)}}\left(\langle\theta_{1}^{(2)},\theta_{1}^{(2)}|\mathcal{O}|\theta_{1}^{(2)},\theta_{1}^{(2)}\rangle_{L}-\langle\mathcal{O}\rangle_{L}\right)\]
for which one needs to evaluate the density of states for a degenerate
two-particle state. The appropriate Bethe-Yang equation reads\begin{equation}
2mL\sinh\theta_{1}^{(2)}+\delta(0)+\delta\left(2\theta_{1}^{(2)}\right)+\delta^{(\alpha)}\left(\theta_{1}^{(2)}\right)+\delta^{(\beta)}\left(\theta_{1}^{(2)}\right)=2\pi I_{1}\label{eq:deg2ptbye}\end{equation}
and so the summation can be replaced by\begin{eqnarray}
\sum_{\theta_{1}^{(2)}=\theta_{2}^{(2)}} & \rightarrow & \int\frac{d\theta}{2\pi}\bar{\rho}_{12}(\theta)\nonumber \\
 &  & \bar{\rho}_{12}(\theta)=2mL\cosh\theta+2\varphi(2\theta)+\psi^{(\alpha)}(\theta_{1})+\psi^{(\beta)}(\theta_{1})\label{eq:degtp}\end{eqnarray}
On the other hand from (\ref{eq:Fcextended}) it follows that\begin{eqnarray*}
\langle\theta,\theta|\mathcal{O}|\theta,\theta\rangle_{L}-\langle\mathcal{O}\rangle_{L} & = & \frac{1}{\rho_{2}(\theta,\theta)}\Big[F_{4}^{c}(\theta_{1},\theta_{2})+\tilde{\rho}(\theta_{1},\theta_{2}|\{1\})F_{2}^{c}(\theta_{1})\\
 &  & +\tilde{\rho}(\theta_{1},\theta_{2}|\{2\})F_{2}^{c}(\theta_{2})\Big]_{\theta_{1}=\theta_{2}=\theta}+O\left(\mathrm{e}^{-\mu L}\right)\end{eqnarray*}
where the $\tilde{\rho}$ are given in (\ref{eq:rhotilde2_1}) and
the factor at the front can be calculated from (\ref{eq:bydet})\[
\rho_{2}(\theta,\theta)=4m^{2}L^{2}\cosh^{2}\theta+O(L)\]
In addition, the exclusion property can be used to substitute $F_{c}^{4}(\theta,\theta)=0$.
Taking the limit $L\rightarrow\infty$ results in\[
\Sigma_{2}^{(1)}=-\frac{1}{2}\int\frac{d\theta}{2\pi}\mathrm{e}^{-2mR\cosh\theta}2F_{2}^{c}(\theta)\]
and so the total contribution at this order reads\begin{eqnarray}
\Sigma_{2} & = & \Sigma_{2}^{(1)}+\Sigma_{2}^{(2)}\nonumber \\
 & = & -\int_{0}^{\infty}\frac{d\theta}{2\pi}\mathrm{e}^{-2mR\cosh\theta}F_{2}^{c}(\theta)\nonumber \\
 & + & \frac{1}{2}\int_{0}^{\infty}\frac{d\theta_{1}}{2\pi}\int_{0}^{\infty}\frac{d\theta_{2}}{2\pi}\mathrm{e}^{-mR(\cosh\theta_{1}+\cosh\theta_{2})}\Big[F_{4}^{c}(\theta_{1},\theta_{2})\nonumber \\
 &  & +\left(\varphi(\theta_{1}-\theta_{2})+\varphi(\theta_{1}+\theta_{2})\right)\left(F_{2}^{c}(\theta_{1})+F_{2}^{c}(\theta_{2})\right)\Big]\label{eq:bff2order}\end{eqnarray}

\subsection{Corrections of order $\mathrm{e}^{-3mR}$}

This calculation proceeds in a similar way but it is rather long and
so it is relegated to appendix C. The net result is

\begin{eqnarray}
\Sigma_{3} & = & \frac{1}{6}\int_{0}^{\infty}\frac{d\theta_{1}}{2\pi}\int_{0}^{\infty}\frac{d\theta_{2}}{2\pi}\int_{0}^{\infty}\frac{d\theta_{3}}{2\pi}\mathrm{e}^{-mR(\cosh\theta_{1}+\cosh\theta_{2}+\cosh\theta_{3})}[F_{6}^{c}(\theta_{1},\theta_{2},\theta_{3})\nonumber \\
 &  & +3F_{4}^{c}(\theta_{2},\theta_{3})(\Phi_{12}+\Phi_{13})+3F_{2}^{c}(\theta_{3})(\Phi_{12}\Phi_{13}+\Phi_{12}\Phi_{23}+\Phi_{13}\Phi_{23})]\nonumber \\
 &  & -\int_{0}^{\infty}\frac{d\theta_{1}}{2\pi}\int_{0}^{\infty}\frac{d\theta_{2}}{2\pi}\mathrm{e}^{-mR(2\cosh\theta_{1}+\cosh\theta_{2})}\Big[F_{4}^{c}(\theta_{1},\theta_{2})\nonumber \\
 &  & +\left(2F_{2}^{c}(\theta_{1})+\frac{1}{2}F_{2}^{c}(\theta_{2})\right)\Phi_{12}\Big]+\int_{0}^{\infty}\frac{d\theta_{1}}{2\pi}\mathrm{e}^{-3mR\cosh\theta_{1}}F_{2}^{c}(\theta_{1})\label{eq:bff3order}\end{eqnarray}
where $\Phi_{ij}=\varphi(\theta_{i}-\theta_{j})+\varphi(\theta_{i}+\theta_{j})$.

\subsection{Discussion of the results}

It is very important to note that in the order by order corrections
(\ref{eq:bff1order}), (\ref{eq:bff2order}) and (\ref{eq:bff3order}),
the dependence on the boundary condition $\beta$ at $x=0$ is only
carried by the form factors $F_{2n}^{c}$. However, in the intermediate
calculations the Bethe-Yang determinants enter, which depend on the
boundary conditions $\alpha$ and $\beta$ in a symmetrical way: according
to eqn. (\ref{eq:jacmatexplicit}), the boundary phase-shift derivatives
always appear in the combination\[
\psi^{(\alpha)}(\theta)+\psi^{(\beta)}(\theta)\]
The fact that all such terms drop in the $L\rightarrow\infty$ limit
is necessary for consistency since the end result can only depend
on the boundary condition $\beta$ imposed at $x=0$, but not on the
auxiliary (and indeed arbitrary) boundary condition $\alpha$ imposed
at $x=-L$ (cf. figure \ref{fig:finvolcylinder}).

Summarizing the results, the expansion of the one-point function reads 

\[
\langle\mathcal{O}\rangle^{R}=\langle\mathcal{O}\rangle+\Sigma_{1}+\Sigma_{2}+\Sigma_{3}+O\left(\mathrm{e}^{-4mR}\right)\]
where \[
\Sigma_{1}=\int_{0}^{\infty}\frac{d\theta}{2\pi}F_{2}^{c}(\theta)\mathrm{e}^{-mR\cosh\theta}\]
while $\Sigma_{2}$ and $\Sigma_{3}$ are given in eqns. (\ref{eq:bff2order})
and (\ref{eq:bff3order}), respectively. Note that this result exactly
coincides with the expansion (\ref{eq:myconjexpanded}) of the conjectured
formula (\ref{eq:myconjecture}), which is a strong reason to believe
that the conjecture is indeed correct to all orders (especially in
view of the very nontrivial structure of the third-order correction
terms).

There is a rather obvious structural similarity between the bulk formula
(\ref{eq:leclmuss1pt}) and the boundary one (\ref{eq:myconjecture}).
Taking into account the symmetry of the pseudo-energy function exploited
in appendix A, it is possible to bring the bulk and boundary cases
into correspondence by interchanging the following ingredients: 

\begin{center}
\begin{tabular}{ccc}
bulk &  & boundary\tabularnewline
$\int_{-\infty}^{\infty}\frac{d\theta_{i}}{2\pi}$ &  & $\int_{0}^{\infty}\frac{d\theta_{i}}{2\pi}$\tabularnewline
$f_{2n}^{c}(\theta_{1},\dots\theta_{n})$ &  & $F_{2n}^{c}(\theta_{1},\dots\theta_{n})$\tabularnewline
$\varphi(\theta_{j}-\theta_{k})$ &  & $\varphi(\theta_{j}-\theta_{k})+\varphi(\theta_{j}+\theta_{k})$\tabularnewline
\end{tabular}
\par\end{center}

\noindent (some care must be taken on the second line to follow properly
the particle labels of $f_{2}^{c}$, since the bulk connected two-particle
form factor is actually independent of the rapidity and thus the argument
is usually omitted). Since Theorem 1 of appendix B is related to the
corresponding bulk theorem of \cite{fftcsa2} via the correspondence
implied by the last two lines in the above table, it is also possible
to express the expansion (\ref{eq:myconjexpanded}) in terms of symmetric
form factors analogously to the result obtained in \cite{fftcsa2}:\begin{eqnarray*}
\langle\mathcal{O}\rangle^{R} & = & \langle\mathcal{O}\rangle+\int_{0}^{\infty}\frac{d\theta}{2\pi}F_{2}^{s}(\theta)\left[\mathrm{e}^{-mR\cosh\theta}-\mathrm{e}^{-2mR\cosh\theta}+\mathrm{e}^{-3mR\cosh\theta}\right]\\
 &  & +\frac{1}{2}\int_{0}^{\infty}\frac{d\theta_{1}}{2\pi}\int_{0}^{\infty}\frac{d\theta_{2}}{2\pi}F_{4}^{s}(\theta_{1},\theta_{2})\left[\mbox{e}^{-mR(\cosh\theta_{1}+\cosh\theta_{2})}-2\mbox{e}^{-mR(2\cosh\theta_{1}+\cosh\theta_{2})}\right]\\
 &  & +\frac{1}{6}\int_{0}^{\infty}\frac{d\theta_{1}}{2\pi}\int_{0}^{\infty}\frac{d\theta_{2}}{2\pi}\int_{0}^{\infty}\frac{d\theta_{3}}{2\pi}F_{6}^{s}(\theta_{1},\theta_{2},\theta_{3})\mbox{e}^{-mR(\cosh\theta_{1}+\cosh\theta_{2}+\cosh\theta_{3})}\\
 &  & -\int_{0}^{\infty}\frac{d\theta_{1}}{2\pi}\int_{0}^{\infty}\frac{d\theta_{2}}{2\pi}\left[F_{2}^{s}(\theta_{1})-\frac{1}{2}F_{2}^{s}(\theta_{2})\right]\Phi_{12}\mbox{e}^{-mR(2\cosh\theta_{1}+\cosh\theta_{2})}+O\left(\mathrm{e}^{-4mR}\right)\end{eqnarray*}
where eqns. (\ref{eq:Fs2},\ref{eq:Fs4},\ref{eq:Fs6}) were used,
together with the freedom to relabel some integration variables. However,
note that this is not automatically guaranteed in the finite volume
formalism used in the present section, since the computation makes
use of the various Bethe-Yang determinants which depend explicitly
on the combination $\varphi(\theta_{j}-\theta_{k})-\varphi(\theta_{j}+\theta_{k})$
as pointed out in appendix B. The agreement between (\ref{eq:myconjexpanded})
and the corrections in eqns. (\ref{eq:bff1order},\ref{eq:bff2order},\ref{eq:bff3order})
shows that this dependence drops out after the limit $L\,\rightarrow\,\infty$,
which is far from trivial, albeit required for overall consistency.

\section{Conclusions and outlook}

The main result of this paper is eqn. (\ref{eq:myconjecture}) (or
its generalization (\ref{eq:genconjecture})) which provides a way
to evaluate finite temperature expectation values of boundary operators
in terms of form factors. 

At first sight all the rest of the argument (i.e. the low-temperature
expansion using the finite volume regularization) is only developed
in order to verify this conjecture. However, as already pointed out
for the bulk case discussed in \cite{fftcsa2}, the finite volume
regulator can be used to evaluate two-point (or even higher) correlation
functions at finite temperature. There has been some development in
the bulk case \cite{leclairmussardo,saleurfiniteT,castrofring,esslerfiniteT,tsvelikfiniteTcorr},
but there is a general problem that the regulator imposed to deal
with the disconnected contributions is rather ad hoc. The failure
of Delfino's proposal for the bulk finite temperature expectation
values \cite{delfinofiniteT,mussardodifference} shows that the ambiguity
inherent in the regularization procedure (which is manifested in the
directional dependence of the diagonal limit discussed in subsection
4.2 and appendix B) must be taken seriously.

However, as pointed out already in \cite{fftcsa2}, finite volume
as a regulator is guaranteed to give a correct answer as a matter
of principle, since it provides a physical way to regularize the form
factors entering the expansion. Therefore it would be very interesting
to apply the ideas presented in \cite{fftcsa2} and here to compute
bulk and boundary two-point functions, respectively.

Another interesting issue is to obtain an extension of the finite
volume description of form factors to non-diagonal scattering theories,
both in the bulk and on the boundary. Since the description of finite
volume energy levels is known and is not very complicated (one obtains
scalar Bethe-Yang equations after suitably diagonalizing a family
of commuting transfer matrices, cf. \cite{nondiag} and references
therein), it can be expected that the necessary description of form
factors is not too difficult to find. One can then use these results
to evaluate finite temperature averages and correlators in the non-diagonal
case as well.

\subsection*{Acknowledgments}

The author is grateful to L. Palla for comments on the manuscript.
This research was partially supported by the Hungarian research fund
OTKA K60040. The author was also supported by a Bolyai J\'anos research
scholarship.

\appendix
\makeatletter 
\renewcommand{\theequation}{\hbox{\normalsize\Alph{section}.\arabic{equation}}} 
\@addtoreset{equation}{section} 
\renewcommand{\thefigure}{\hbox{\normalsize\Alph{section}.\arabic{figure}}} 
\@addtoreset{figure}{section} 
\renewcommand{\thetable}{\hbox{\normalsize\Alph{section}.\arabic{table}}} 
\@addtoreset{table}{section} 
\makeatother

\section{Low-temperature expansion of the conjectured formula (\ref{eq:leclmuss1pt})}

First the pseudo-energy function $\epsilon(\theta)$ must be expanded
to the necessary order. Using the fact that $\epsilon(\theta)$ is
an even function, the TBA equation can be written in the form \[
\epsilon(\theta)=mR\cosh(\theta)-\int_{0}^{\infty}\frac{d\theta'}{2\pi}\left[\varphi(\theta-\theta')+\varphi(\theta+\theta')\right]\log(1+\mathrm{e}^{-\epsilon(\theta')})\]
Iterating this equation twice with the starting value $\epsilon^{(0)}(\theta)=mR\cosh(\theta)$
and taking care to expand the logarithm one obtains\begin{eqnarray*}
\epsilon(\theta_{1}) & = & RE_{1}-\int_{0}^{\infty}\frac{d\theta_{2}}{2\pi}\Phi_{12}\mathrm{e}^{-RE_{2}}-\frac{1}{2}\int_{0}^{\infty}\frac{d\theta_{2}}{2\pi}\Phi_{12}\mathrm{e}^{-2RE_{2}}\\
 & - & \int_{0}^{\infty}\frac{d\theta_{2}}{2\pi}\int_{0}^{\infty}\frac{d\theta_{3}}{2\pi}\Phi_{12}\Phi_{23}\mathrm{e}^{-R(E_{2}+E_{3})}+O\left(\mathrm{e}^{-3mR}\right)\end{eqnarray*}
where\[
E_{i}=m\cosh\theta_{i}\quad,\quad\Phi_{ij}=\varphi(\theta_{i}-\theta_{j})+\varphi(\theta_{i}+\theta_{j})\]
which leads to \begin{eqnarray}
\mathrm{e}^{-\epsilon(\theta_{1})} & = & \mathrm{e}^{-RE_{1}}+\mathrm{e}^{-RE_{1}}\int_{0}^{\infty}\frac{d\theta_{2}}{2\pi}\Phi_{12}\mathrm{e}^{-RE_{2}}\nonumber \\
 &  & +\frac{1}{2}\mathrm{e}^{-RE_{1}}\left(\int_{0}^{\infty}\frac{d\theta_{2}}{2\pi}\Phi_{12}\mathrm{e}^{-RE_{2}}\right)^{2}-\frac{1}{2}\mathrm{e}^{-RE_{1}}\int_{0}^{\infty}\frac{d\theta_{2}}{2\pi}\Phi_{12}\mathrm{e}^{-2RE_{2}}\nonumber \\
 &  & +\mathrm{e}^{-RE_{1}}\int_{0}^{\infty}\frac{d\theta_{2}}{2\pi}\int_{0}^{\infty}\frac{d\theta_{3}}{2\pi}\Phi_{12}\Phi_{23}\mathrm{e}^{-R(E_{2}+E_{3})}+O\left(\mathrm{e}^{-4mR}\right)\label{eq:psenexpanded}\end{eqnarray}
Recall that (\ref{eq:myconjecture}) reads \[
\langle\mathcal{O}\rangle^{R}=\sum_{n=0}^{\infty}\frac{1}{n!}\prod_{i=1}^{n}\left(\int_{0}^{\infty}\frac{d\theta_{i}}{2\pi}\frac{\mathrm{e}^{-\epsilon(\theta_{i})}}{1+\mathrm{e}^{-\epsilon(\theta_{i})}}\right)F_{2n}^{c}(\theta_{1},...,\theta_{n})\]
Using (\ref{eq:psenexpanded}) and the geometric series \[
\frac{\mathrm{e}^{-\epsilon}}{1+\mathrm{e}^{-\epsilon}}=\mathrm{e}^{-\epsilon}-\mathrm{e}^{-2\epsilon}+\mathrm{e}^{-3\epsilon}+\dots\]
this can be expanded in orders of $\mbox{e}^{-mR}$. One obtains\begin{equation}
\langle\mathcal{O}\rangle^{R}=\sigma_{1}+\sigma_{2}+\sigma_{3}+O\left(\mbox{e}^{-4mR}\right)\label{eq:conj_expanded_appendix}\end{equation}
where\begin{eqnarray}
\sigma_{1} & = & \int_{0}^{\infty}\frac{d\theta_{1}}{2\pi}\left(\mathrm{e}^{-RE_{1}}-\mathrm{e}^{-2RE_{1}}+\mathrm{e}^{-3RE_{1}}\right)F_{2}^{c}(\theta_{1})+\int_{0}^{\infty}\frac{d\theta_{1}}{2\pi}\int_{0}^{\infty}\frac{d\theta_{2}}{2\pi}\mathrm{e}^{-R(E_{1}+E_{2})}\Phi_{12}F_{2}^{c}(\theta_{1})\nonumber \\
 &  & +\frac{1}{2}\int_{0}^{\infty}\frac{d\theta_{1}}{2\pi}\int_{0}^{\infty}\frac{d\theta_{2}}{2\pi}\int_{0}^{\infty}\frac{d\theta_{3}}{2\pi}\mathrm{e}^{-R(E_{1}+E_{2}+E_{3})}\left(\Phi_{12}\Phi_{13}+\Phi_{12}\Phi_{23}+\Phi_{13}\Phi_{23}\right)F_{2}^{c}(\theta_{1})\nonumber \\
 &  & -\int_{0}^{\infty}\frac{d\theta_{1}}{2\pi}\int_{0}^{\infty}\frac{d\theta_{2}}{2\pi}\left(2\mathrm{e}^{-R(2E_{1}+E_{2})}+\frac{1}{2}\mathrm{e}^{-R(E_{1}+2E_{2})}\right)\Phi_{12}F_{2}^{c}(\theta_{1})\nonumber \\
\sigma_{2} & = & \frac{1}{2}\int_{0}^{\infty}\frac{d\theta_{1}}{2\pi}\int_{0}^{\infty}\frac{d\theta_{2}}{2\pi}\mathrm{e}^{-R(E_{1}+E_{2})}F_{4}^{c}(\theta_{1},\theta_{2})\nonumber \\
 &  & -\int_{0}^{\infty}\frac{d\theta_{1}}{2\pi}\int_{0}^{\infty}\frac{d\theta_{2}}{2\pi}\mathrm{e}^{-R(2E_{1}+E_{2})}F_{4}^{c}(\theta_{1},\theta_{2})\nonumber \\
 &  & +\frac{1}{2}\int_{0}^{\infty}\frac{d\theta_{1}}{2\pi}\int_{0}^{\infty}\frac{d\theta_{2}}{2\pi}\int_{0}^{\infty}\frac{d\theta_{3}}{2\pi}\mathrm{e}^{-R(E_{1}+E_{2}+E_{3})}\left(\Phi_{13}+\Phi_{23}\right)F_{4}^{c}(\theta_{1},\theta_{2})\nonumber \\
\sigma_{3} & = & \frac{1}{6}\int_{0}^{\infty}\frac{d\theta_{1}}{2\pi}\int_{0}^{\infty}\frac{d\theta_{2}}{2\pi}\int_{0}^{\infty}\frac{d\theta_{3}}{2\pi}\mathrm{e}^{-R(E_{1}+E_{2}+E_{3})}F_{6}^{c}(\theta_{1},\theta_{2},\theta_{3})\label{eq:smallsigmas_appendix}\end{eqnarray}
are the one/two/three-particle contributions expanded to $O(\mathrm{e}^{-4mR})$.

\section{Relation between different evaluations of the diagonal matrix element}

Here the arguments of \cite{fftcsa2} are generalized to the case
of boundary form factors. The goal is to compute the general expression
\begin{equation}
F_{a_{1}\dots a_{n}}(\theta_{1},\dots,\theta_{n}|\epsilon_{1},\dots,\epsilon_{n})=F_{\bar{a}_{n}\dots\bar{a}_{1}a_{1}\dots a_{n}}(\theta_{n}+i\pi+\epsilon_{n},...,\theta_{1}+i\pi+\epsilon_{1},\theta_{1},...,\theta_{n})\label{f2n_eztkellene}\end{equation}
for infinitesimal values of the $\epsilon_{i}$. It is also interesting
to consider the symmetric evaluation

\begin{equation}
F_{a_{1}\dots a_{n}}^{s}(\theta_{1},\dots,\theta_{n})=\lim_{\epsilon\rightarrow0}F_{\bar{a}_{n}\dots\bar{a}_{1}a_{1}\dots a_{n}}(\theta_{n}+i\pi+\epsilon,...,\theta_{1}+i\pi+\epsilon,\theta_{1},...,\theta_{n})\label{eq:symmetric_def}\end{equation}
Let us take $n$ vertices labeled by the numbers $1,2,\dots,n$ and
let $G$ be the set of the directed graphs $G_{i}$ with the following
properties: 

\begin{itemize}
\item $G_{i}$ is tree-like. 
\item For each vertex there is at most one outgoing edge. 
\end{itemize}
For an edge going from $i$ to $j$ we use the notation $E_{ij}$.

\paragraph{Theorem 1\label{par:Theorem-1}}

(\ref{f2n_eztkellene}) can be evaluated as a sum over all graphs
in $G$, where the contribution of a graph $G_{i}$ is given by the
following two rules: 

\begin{itemize}
\item Let $A_{i}=\{\alpha_{1},\alpha_{2},\dots,\alpha_{m}\}$ be the set
of vertices from which there are no outgoing edges in $G_{i}$. The
form factor associated to $G_{i}$ is \begin{equation}
F_{a_{\alpha_{1}}\dots a_{\alpha_{m}}}^{c}(\theta_{a_{1}},\theta_{a_{2}},\dots,\theta_{a_{m}})\label{egygrafformfaktora}\end{equation}
 
\item For each edge $E_{jk}$ the form factor above has to be multiplied
by \[
\frac{\epsilon_{j}}{\epsilon_{k}}\Phi_{jk}\]
where\[
\Phi_{jk}=\varphi_{a_{j}a_{k}}(\theta_{j}-\theta_{k})+\varphi_{a_{j}a_{k}}(\theta_{j}+\theta_{k})=\Phi_{kj}\]
 
\end{itemize}

\paragraph*{Proof}

The proof goes by induction in $n$. For $n=1$ there is only a single
way to take the limit and so \[
F_{a}(\theta_{1}|\epsilon_{1})=F_{a}^{c}(\theta_{1})=F_{\bar{a}a}(i\pi+\theta_{1},\theta_{1})\]
This is in accordance with the theorem, because for $n=1$ there is
only the trivial graph which contains no edges and a single node. 

Now assume that the theorem is true for $n-1$ and let us take the
case of $n$ particles. Consider the residue of the matrix element
(\ref{f2n_eztkellene}) at $\epsilon_{n}=0$ while keeping all the
$\epsilon_{i}$ finite \[
R=\mathop{\mathrm{Res}}_{\epsilon_{n}=0}F_{a_{1}\dots a_{n}}(\theta_{1}..\theta_{n}|\epsilon_{1}..\epsilon_{n})\]
According to the theorem the graphs contributing to this residue are
exactly those for which the vertex $n$ has an outgoing edge and no
incoming edges. Let $R_{j}$ be sum of the diagrams where the outgoing
edge is $E_{nj}$ for some $j=1,\dots,n-1$, and so \[
R=\sum_{j=1}^{n-1}R_{j}\]
The form factors appearing in $R_{j}$ do not depend on $\theta_{n}$.
Therefore one gets exactly the diagrams that are needed to evaluate
$F_{2(n-1)}(\theta_{1}..\theta_{n-1}|\epsilon_{1}..\epsilon_{n-1})$,
apart from the proportionality factor associated to the link $E_{nj}$
and so \[
R_{j}=\epsilon_{j}\Phi_{jn}F_{a_{1}\dots a_{n-1}}(\theta_{1}..\theta_{n-1}|\epsilon_{1}..\epsilon_{n-1})\]
and summing over $j$ yields \begin{equation}
R=(\epsilon_{1}\Phi_{1n}+\epsilon_{2}\Phi_{2n}+\dots+\epsilon_{n-1}\Phi_{n-1\, n})F_{a_{1}\dots a_{n-1}}(\theta_{1}..\theta_{n-1}|\epsilon_{1}..\epsilon_{n-1})\label{erremitlepsz}\end{equation}
In order to prove the theorem, one only needs to show that the residue
indeed takes this form. On the other hand, using the kinematical residue
axiom (\ref{eq:kinematical}) \begin{eqnarray*}
R & = & i\left(1-\prod_{j=1}^{n-1}S_{a_{n}a_{j}}(\theta_{n}-\theta_{j})S_{a_{n}a_{j}}(\theta_{n}-\theta_{j}-i\pi-\epsilon_{j})S_{a_{n}a_{j}}(\theta_{n}+\theta_{i})S_{a_{n}a_{j}}(\theta_{n}+\theta_{j}+i\pi+\epsilon_{j})\right)\\
 &  & \times F_{a_{1}\dots a_{n-1}}(\theta_{1}..\theta_{n-1}|\epsilon_{1}..\epsilon_{n-1})\end{eqnarray*}
 which is exactly the same as eqn. (\ref{erremitlepsz}) when expanded
to first order in $\epsilon_{j}$.

Therefore the procedure described in the theorem gives the correct
result for the terms that include a $1/\epsilon_{n}$ singularity.
Using symmetry in the rapidity variables this is true for all the
terms that include at least one $1/\epsilon_{i}$ for an arbitrary
$i$. There is only one diagram that cannot be generated by the inductive
procedure, namely the empty graph. However, there are no singularities
($1/\epsilon_{i}$ factors) associated to it, and it is identical
to $F_{2n}^{c}(\theta_{1},\dots,\theta_{n})$ by definition. \emph{Qed}.

\begin{figure}
\noindent \begin{centering}
\includegraphics[scale=1.2]{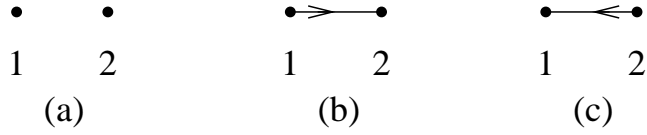}
\par\end{centering}

\caption{\label{fig:neq2gr} The graphs relevant for $n=2$}

\end{figure}

\begin{figure}
\noindent \begin{centering}
\includegraphics[scale=1.2]{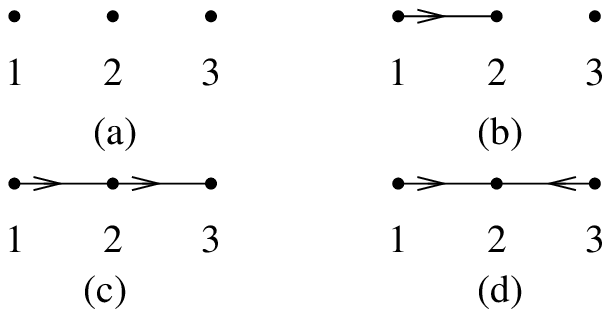}
\par\end{centering}

\caption{\label{fig:neq3gr} The graphs relevant for $n=3$}

\end{figure}

Let us now illustrate how the theorem works for the case of a theory
with a single particle species. In this case one can use the simplified
notation introduced in (\ref{eq:connectedsimplenotation}) and similarly
denote\[
F_{2n}(\theta_{1},\dots,\theta_{n}|\epsilon_{1},\dots,\epsilon_{n})=F(\theta_{n}+i\pi+\epsilon_{n},...,\theta_{1}+i\pi+\epsilon_{1},\theta_{1},...,\theta_{n})\]
The case $n=1$ is trivial:\begin{equation}
F_{2}^{c}(\theta)=F_{2}^{s}(\theta)\label{eq:Fs2}\end{equation}
For $n=2$, there are only three graphs, depicted in figure \ref{fig:neq2gr}.
Applying the rules yields \[
F_{4}(\theta_{1},\theta_{2}|\epsilon_{1},\epsilon_{2})=F_{4}^{c}(\theta_{1},\theta_{2})+\Phi_{12}\left(\frac{\epsilon_{1}}{\epsilon_{2}}F_{2}^{c}(\theta_{2})+\frac{\epsilon_{2}}{\epsilon_{1}}F_{2}^{c}(\theta_{1})\right)\]
which yields \begin{equation}
F_{4}^{s}(\theta_{1},\theta_{2})=F_{4}^{c}(\theta_{1},\theta_{2})+\Phi_{12}\left(F_{2}^{c}(\theta_{2})+F_{2}^{c}(\theta_{1})\right)\label{eq:Fs4}\end{equation}
upon putting $\epsilon_{1}=\epsilon_{2}$. For $n=3$ there are $4$
different kinds of graphs, the representatives of which are shown
in figure \ref{fig:neq3gr}; all other graphs can be obtained by permuting
the node labels $1,2,3$. The contributions of these graphs are

\begin{eqnarray*}
(a) & : & F_{6}^{c}(\theta_{1},\theta_{2},\theta_{3})\\
(b) & : & \frac{\epsilon_{2}}{\epsilon_{1}}\Phi_{12}F_{4}^{c}(\theta_{2},\theta_{3})\\
(c) & : & \frac{\epsilon_{2}}{\epsilon_{1}}\frac{\epsilon_{3}}{\epsilon_{2}}\Phi_{12}\Phi_{23}F_{2}^{c}(\theta_{3})=\frac{\epsilon_{3}}{\epsilon_{1}}\Phi_{12}\Phi_{23}F_{2}^{c}(\theta_{3})\\
(d) & : & \frac{\epsilon_{2}}{\epsilon_{1}}\frac{\epsilon_{2}}{\epsilon_{3}}\Phi_{12}\Phi_{23}F_{2}^{c}(\theta_{2})\end{eqnarray*}
Adding up all the contributions and putting $\epsilon_{1}=\epsilon_{2}=\epsilon_{3}$:
\begin{eqnarray}
F_{6}^{s}(\theta_{1},\theta_{2},\theta_{3}) & = & F_{6}^{c}(\theta_{1},\theta_{2},\theta_{3})\nonumber \\
 & + & (\Phi_{12}+\Phi_{13})F_{4}^{c}(\theta_{2},\theta_{3})+(\Phi_{12}+\Phi_{23})F_{4}^{c}(\theta_{1},\theta_{3})+(\Phi_{13}+\Phi_{23})F_{4}^{c}(\theta_{1},\theta_{2})\nonumber \\
 & + & (F_{2}^{c}(\theta_{1})+F_{2}^{c}(\theta_{2})+F_{2}^{c}(\theta_{3}))(\Phi_{12}\Phi_{13}+\Phi_{12}\Phi_{23}+\Phi_{13}\Phi_{23})\label{eq:Fs6}\end{eqnarray}
It can be seen that these results are a natural generalization of
the bulk ones obtained in \cite{fftcsa2} with $\varphi$ replaced
by $\Phi$. It is also important to keep in mind that contrary to
the bulk situation $F_{2}^{c}$ depends on the rapidity (in the bulk
it is a constant since Lorentz invariance entails that all form factors
depend only on rapidity differences). 

Now the finite volume diagonal matrix elements (\ref{eq:diaggenrulesaleur})
can also be re-expressed in terms of the symmetric evaluation. The
first nontrivial case is $n=2$ for which\begin{eqnarray*}
\langle\{I_{1},I_{2}\}|\mathcal{O}(0)|\{I_{1},I_{2}\}\rangle & = & \frac{1}{\rho_{2}(\tilde{\theta}_{1},\tilde{\theta}_{2})}\Big(F_{4}^{c}(\tilde{\theta}_{1},\tilde{\theta}_{2})+\tilde{\rho}(\tilde{\theta}_{1},\tilde{\theta}_{2}|\{1\})F_{2}^{c}(\tilde{\theta}_{1})\\
 &  & +\tilde{\rho}(\tilde{\theta}_{1},\tilde{\theta}_{2}|\{2\})F_{2}^{c}(\tilde{\theta}_{2})\Big)+\left\langle \mathcal{O}\right\rangle +O(\mathrm{e}^{-\mu L})\end{eqnarray*}
where $\tilde{\theta}_{1},\tilde{\theta}_{2}$ are the solutions of
the $2$-particle Bethe-Yang equations with quantum numbers $I_{1},I_{2}$.
$\tilde{\rho}$ denotes the appropriate sub-determinants (\ref{eq:jacsubmat})
of the two-particle Jacobian matrix, while $\rho_{n}$ is the full
$n$-particle Jacobi determinant (\ref{eq:bydet}). It is straightforward
to verify that\begin{eqnarray}
F_{4}^{c}(\theta_{1},\theta_{2})+\tilde{\rho}(\theta_{1},\theta_{2}|\{1\})F_{2}^{c}(\theta_{1})+\tilde{\rho}(\theta_{1},\theta_{2}|\{2\})F_{2}^{c}(\theta_{2})=\nonumber \\
F_{4}^{s}(\theta_{1},\theta_{2})+\rho_{1}(\theta_{1})F_{2}^{s}(\theta_{1})+\rho_{1}(\theta_{2})F_{2}^{s}(\theta_{2})\nonumber \\
+2\varphi(\theta_{1}+\theta_{2})\left(F_{2}^{s}(\theta_{1})+F_{2}^{s}(\theta_{2})\right)\label{eq:th2fails}\end{eqnarray}
The term on the last line shows that the analogue of Theorem 2 in
\cite{fftcsa2} (which would make the expressions on the first two
lines identical) fails in the boundary case. This results from the
fact that the derivative matrix (\ref{eq:jacmatexplicit}) of the
Bethe-Yang equations (\ref{eq:bbye}) carries a dependence not only
on the combination $\varphi(\theta_{j}-\theta_{k})+\varphi(\theta_{j}+\theta_{k})$,
but also on $\varphi(\theta_{j}-\theta_{k})-\varphi(\theta_{j}+\theta_{k})$.

The relations (\ref{eq:Fs4}), (\ref{eq:Fs6}) and (\ref{eq:th2fails})
were also verified numerically using the explicit form factor solutions
presented in \cite{bfftcsa}.

\section{$\mathrm{e}^{-3mR}$ corrections to the finite temperature one-point
function}

In order to keep the calculation manageable, let us introduce the
following shortened notations:\begin{eqnarray*}
 &  & E_{i}=m\cosh\theta_{i}\\
 &  & \langle\theta_{1},\dots,\theta_{n}|\mathcal{O}|\theta_{1},\dots,\theta_{n}\rangle_{L}=\langle1\dots n|\mathcal{O}|1\dots n\rangle_{L}\\
 &  & \rho(\theta_{1},\dots,\theta_{n})=\rho(1\dots n)\\
 &  & \tilde{\rho}(\theta_{1},\dots,\theta_{n}|\{a_{1},\dots,a_{k}\})=\tilde{\rho}(1\dots n|\{a_{1},\dots,a_{k}\})\end{eqnarray*}
Summations will be shortened to\begin{eqnarray*}
\sum_{\theta_{1}\dots\theta_{n}} & \rightarrow & \sum_{1\dots n}\\
\sum_{\theta_{1}\dots\theta_{n}}{}^{'} & \rightarrow & \sum_{1\dots n}{}^{'}\end{eqnarray*}
and for later convenience also denote\[
\Phi_{ij}=\varphi(\theta_{i}-\theta_{j})+\varphi(\theta_{i}+\theta_{j})\]
which satisfies $\Phi_{ij}=\Phi_{ji}$. 

Multiplying (\ref{eq:nomexp}) with (\ref{eq:Zinvexp}) and collecting
the third order correction terms:\begin{eqnarray*}
 &  & \frac{1}{6}\sum_{123}{}^{'}\mathrm{e}^{-R(E_{1}+E_{2}+E_{3})}\left(\langle123|\mathcal{O}|123\rangle_{L}-\langle\mathcal{O}\rangle_{L}\right)\\
 & - & \left(\sum_{1}\mathrm{e}^{-RE_{1}}\right)\frac{1}{2}\sum_{23}{}^{'}\mathrm{e}^{-R(E_{2}+E_{3})}\left(\langle23|\mathcal{O}|23\rangle_{L}-\langle\mathcal{O}\rangle_{L}\right)\\
 & + & \left\{ \left(\sum_{1}\mathrm{e}^{-RE_{1}}\right)\left(\sum_{2}\mathrm{e}^{-RE_{2}}\right)-\frac{1}{2}\sum_{12}{}^{'}\mathrm{e}^{-R(E_{1}+E_{2})}\right\} \left(\sum_{3}\mathrm{e}^{-RE_{3}}\right)\left(\langle3|\mathcal{O}|3\rangle_{L}-\langle\mathcal{O}\rangle_{L}\right)\end{eqnarray*}
To keep trace of the state densities it is important to avoid combining
rapidity sums. The constrained summations can be replaced by free
sums with the diagonal contributions subtracted:\begin{eqnarray*}
\sum_{12}{}^{'} & = & \sum_{12}-\sum_{1=2}\\
\sum_{123}{}^{'} & = & \sum_{123}-\left(\sum_{1=2,3}+\sum_{2=3,1}+\sum_{1=3,2}\right)+2\sum_{1=2=3}\end{eqnarray*}
where the diagonal contributions are labeled according to which diagonal
the summation corresponds to, but otherwise the given sum is free,
e.g.\[
\sum_{1=2,3}\]
shows a summation over all triplets $\theta_{1}^{(3)},\theta_{2}^{(3)},\theta_{3}^{(3)}$
where $\theta_{1}^{(3)}=\theta_{2}^{(3)}$ and $\theta_{3}^{(3)}$
runs free (it can also be equal with the other two). Finally denote
\[
F(12\dots n)=F_{2n}^{c}(\theta_{1},\dots,\theta_{n})\]
so from (\ref{eq:Fcextended}) the necessary matrix elements can be
written in the form\begin{eqnarray}
\rho(123)\left(\langle123|\mathcal{O}|123\rangle_{L}-\langle\mathcal{O}\rangle_{L}\right) & = & F(123)+\tilde{\rho}(123|\{1,2\})F(12)\nonumber \\
 &  & +\tilde{\rho}(123|\{1,3\})F(13)+\tilde{\rho}(123|\{2,3\})F(23)\nonumber \\
 &  & +\tilde{\rho}(123|\{1\})F(1)+\tilde{\rho}(123|\{2\})F(2)+\tilde{\rho}(123|\{3\})F(3)\nonumber \\
\rho(122)\left(\langle122|\mathcal{O}|122\rangle_{L}-\langle\mathcal{O}\rangle_{L}\right) & = & 2\tilde{\rho}(122|\{1,2\})F(12)+\tilde{\rho}(122|\{1\})F(1)+2\tilde{\rho}(122|\{2\})F(2)\nonumber \\
\rho(111)\left(\langle111|\mathcal{O}|111\rangle_{L}-\langle\mathcal{O}\rangle_{L}\right) & = & 3\tilde{\rho}(111|\{1\})F(1)\nonumber \\
\rho(12)\left(\langle12|\mathcal{O}|12\rangle_{L}-\langle\mathcal{O}\rangle_{L}\right) & = & F(12)+\tilde{\rho}(12|\{1\})F(1)+\tilde{\rho}(12|\{2\})F(2)\nonumber \\
\rho(11)\left(\langle11|\mathcal{O}|11\rangle_{L}-\langle\mathcal{O}\rangle_{L}\right) & = & 2\tilde{\rho}(11|\{1\})F(1)\nonumber \\
\rho(1)\left(\langle1|\mathcal{O}|1\rangle_{L}-\langle\mathcal{O}\rangle_{L}\right) & = & F(1)\label{eq:matelms}\end{eqnarray}
where the exclusion property was already used to eliminate form factors
with equal rapidity arguments.

One can now proceed by collecting terms according to the number of
free rapidity variables. The terms containing threefold summation
are\begin{eqnarray*}
\Sigma_{3}^{(3)} & = & \frac{1}{6}\sum_{123}\mathrm{e}^{-R(E_{1}+E_{2}+E_{3})}\left(\langle123|\mathcal{O}|123\rangle_{L}-\langle\mathcal{O}\rangle_{L}\right)-\frac{1}{2}\sum_{1}\sum_{2,3}\left(\langle23|\mathcal{O}|23\rangle_{L}-\langle\mathcal{O}\rangle_{L}\right)\\
 & + & \left(\sum_{1}\sum_{2}\sum_{3}-\frac{1}{2}\sum_{1,2}\sum_{3}\right)\left(\langle3|\mathcal{O}|3\rangle_{L}-\langle\mathcal{O}\rangle_{L}\right)\end{eqnarray*}
Replacing the sums with integrals\begin{eqnarray*}
\sum_{1} & \rightarrow & \int\frac{d\theta_{1}}{2\pi}\rho(1)\\
\sum_{1,2} & \rightarrow & \int\frac{d\theta_{1}}{2\pi}\frac{d\theta_{2}}{2\pi}\rho(12)\\
\sum_{1,2,3} & \rightarrow & \int\frac{d\theta_{1}}{2\pi}\frac{d\theta_{2}}{2\pi}\frac{d\theta_{3}}{2\pi}\rho(123)\end{eqnarray*}
and using (\ref{eq:matelms})\begin{eqnarray*}
\Sigma_{3}^{(3)} & = & \frac{1}{6}\int\frac{d\theta_{1}}{2\pi}\frac{d\theta_{2}}{2\pi}\frac{d\theta_{3}}{2\pi}\mathrm{e}^{-R(E_{1}+E_{2}+E_{3})}\left(F(123)+3\tilde{\rho}(123|\{2,3\})F(23)+3\tilde{\rho}(123|\{3\})F(3)\right)\\
 & - & \frac{1}{2}\int\frac{d\theta_{1}}{2\pi}\frac{d\theta_{2}}{2\pi}\frac{d\theta_{3}}{2\pi}\mathrm{e}^{-R(E_{1}+E_{2}+E_{3})}\rho(1)\left(F(23)+2\rho(23|\{3\})F(3)\right)\\
 & + & \int\frac{d\theta_{1}}{2\pi}\frac{d\theta_{2}}{2\pi}\frac{d\theta_{3}}{2\pi}\mathrm{e}^{-R(E_{1}+E_{2}+E_{3})}\left(\rho(1)\rho(2)-\frac{1}{2}\rho(12)\right)F(3)\end{eqnarray*}
where some of the integration variables were reshuffled. The result
is\begin{eqnarray}
\Sigma_{3}^{(3)} & = & \frac{1}{6}\int_{0}^{\infty}\frac{d\theta_{1}}{2\pi}\int_{0}^{\infty}\frac{d\theta_{2}}{2\pi}\int_{0}^{\infty}\frac{d\theta_{3}}{2\pi}\mathrm{e}^{-mR(\cosh\theta_{1}+\cosh\theta_{2}+\cosh\theta_{3})}\Big[F_{6}^{c}(\theta_{1},\theta_{2},\theta_{3})\nonumber \\
 &  & +3F_{4}^{c}(\theta_{2},\theta_{3})(\Phi_{12}+\Phi_{13})+3F_{2}^{c}(\theta_{3})(\Phi_{12}\Phi_{13}+\Phi_{12}\Phi_{23}+\Phi_{13}\Phi_{23})\Big]\label{eq:res3int}\end{eqnarray}
(to derive the term on the second line note that the $F(3)$ terms
in the integrand of $\Sigma_{3}^{(3)}$ can be symmetrized in $\theta_{1}$
and $\theta_{2}$ without changing the value of the integral).

It is also easy to deal with terms containing a single integral. The
only term of this form is\[
\Sigma_{3}^{(1)}=\frac{1}{3}\sum_{1=2=3}\mathrm{e}^{-R(E_{1}+E_{2}+E_{3})}\left(\langle123|\mathcal{O}|123\rangle_{L}-\langle\mathcal{O}\rangle_{L}\right)\]
When all rapidities $\theta_{1}^{(3)},\theta_{2}^{(3)},\theta_{3}^{(3)}$
are equal, the three-particle Bethe-Yang equations reduce to%
\footnote{Just as in (\ref{eq:deg2ptbye}) there are also contributions of the
form $\delta(0)$, but these can be absorbed into a redefinition of
$I_{1}$.%
} \[
2mL\sinh\theta_{1}^{(3)}+2\delta\left(2\theta_{1}^{(3)}\right)+\delta^{(\alpha)}\left(\theta_{1}^{(3)}\right)+\delta^{(\beta)}\left(\theta_{1}^{(3)}\right)=2\pi I_{1}\]
Therefore the relevant state density is\[
\bar{\rho}_{123}(\theta)=2mL\cosh\theta+4\varphi(2\theta)+\psi^{(\alpha)}(\theta)+\psi^{(\beta)}(\theta)\]
 and\begin{eqnarray}
\Sigma_{3}^{(1)} & = & \frac{1}{3}\int\frac{d\theta_{1}}{2\pi}\mathrm{e}^{-3RE_{1}}\bar{\rho}_{123}(\theta_{1})\left(\langle111|\mathcal{O}|111\rangle_{L}-\langle\mathcal{O}\rangle_{L}\right)\nonumber \\
 & = & \int\frac{d\theta_{1}}{2\pi}\mathrm{e}^{-3RE_{1}}\rho(1)\frac{\tilde{\rho}(111|\{1\})}{\rho(111)}F(1)\,\mathop{\rightarrow}_{L\rightarrow\infty}\,\int\frac{d\theta_{1}}{2\pi}\mathrm{e}^{-3mR\cosh\theta_{1}}F_{2}^{c}(\theta_{1})\label{eq:res1int}\end{eqnarray}
where it was used that\[
\rho(1)\frac{\tilde{\rho}(111|\{1\})}{\rho(111)}\rightarrow1\]
when $L\rightarrow\infty$.

The calculation of double integral terms is much more involved. The
contributions containing two rapidity summations are \begin{eqnarray}
\Sigma_{3}^{(2)} & = & -\frac{1}{6}\left(\sum_{1=2,3}+\sum_{1=3,2}+\sum_{2=3,1}\right)\mathrm{e}^{-R(E_{1}+E_{2}+E_{3})}\left(\langle123|\mathcal{O}|123\rangle_{L}-\langle\mathcal{O}\rangle_{L}\right)\nonumber \\
 &  & +\frac{1}{2}\sum_{1}\sum_{2=3}\mathrm{e}^{-R(E_{1}+E_{2}+E_{3})}\left(\langle23|\mathcal{O}|23\rangle_{L}-\langle\mathcal{O}\rangle_{L}\right)\nonumber \\
 &  & +\frac{1}{2}\sum_{1=2}\sum_{3}\mathrm{e}^{-R(E_{1}+E_{2}+E_{3})}\left(\langle3|\mathcal{O}|3\rangle_{L}-\langle\mathcal{O}\rangle_{L}\right)\label{eq:dintstart}\end{eqnarray}
The density of partially degenerate two-particle states was already
computed in (\ref{eq:degtp}), but the density of partially degenerate
three-particle states is also needed. The relevant Bethe-Yang equations
are%
\footnote{Just as in (\ref{eq:deg2ptbye}) there are also contributions of the
form $\delta(0)$, but these can be absorbed into a redefinition of
$I_{1}$.%
}\begin{eqnarray*}
2mL\sinh\theta_{1}+\delta(\theta_{1}-\theta_{2})+\delta(\theta_{1}+\theta_{2})+\delta(2\theta_{1})+\delta^{(\alpha)}\left(\theta_{1}\right)+\delta^{(\beta)}\left(\theta_{1}\right) & = & 2\pi I_{1}\\
2mL\sinh\theta_{2}+2\delta(\theta_{2}-\theta_{1})+2\delta(\theta_{2}+\theta_{1})+\delta^{(\alpha)}\left(\theta_{2}\right)+\delta^{(\beta)}\left(\theta_{2}\right) & = & 2\pi I_{2}\end{eqnarray*}
where the first and the third particles are put as degenerate (i.e.
$I_{3}=I_{1}$). The density of these degenerate states is then \begin{eqnarray}
\bar{\rho}_{13,2}(12) & = & \det\left(\begin{array}{ll}
r_{11} & r_{12}\\
r_{21} & r_{22}\end{array}\right)\label{eq:p3ptdeg}\\
 &  & r_{11}=2LE_{1}+\varphi(\theta_{1}-\theta_{2})+\varphi(\theta_{1}+\theta_{2})+2\varphi(2\theta_{1})+\psi^{(\alpha)}(\theta)+\psi^{(\beta)}(\theta)\nonumber \\
 &  & r_{22}=2LE_{2}+2\varphi(\theta_{1}-\theta_{2})+2\varphi(\theta_{1}+\theta_{2})+\psi^{(\alpha)}(\theta)+\psi^{(\beta)}(\theta)\nonumber \\
 &  & r_{12}=-\varphi(\theta_{1}-\theta_{2})+\varphi(\theta_{1}+\theta_{2})\quad,\quad r_{21}=-2\varphi(\theta_{1}-\theta_{2})+2\varphi(\theta_{1}+\theta_{2})\nonumber \end{eqnarray}
where it was used that $\varphi(\theta)=\varphi(-\theta)$. Using
the above result and substituting integrals for the sums, eqn. (\ref{eq:dintstart})
can be rewritten in the form\begin{eqnarray*}
 & - & \frac{1}{6}\int\frac{d\theta_{1}}{2\pi}\frac{d\theta_{2}}{2\pi}\mathrm{e}^{-R(2E_{1}+E_{2})}\frac{\bar{\rho}_{13,2}(12)}{\rho(112)}\Big[2\tilde{\rho}(112|\{2,3\})F(12)\\
 &  & +2\tilde{\rho}(112|\{1\})F(1)+\tilde{\rho}(112|\{3\})F(2)+\dots\Big]\\
 & + & \frac{1}{2}\int\frac{d\theta_{1}}{2\pi}\frac{d\theta_{2}}{2\pi}\mathrm{e}^{-R(E_{1}+2E_{2})}\rho(1)\bar{\rho}_{12}(2)\frac{2\tilde{\rho}(22|\{1\})}{\rho(22)}F(2)\\
 & + & \frac{1}{2}\int\frac{d\theta_{1}}{2\pi}\frac{d\theta_{3}}{2\pi}\mathrm{e}^{-R(2E_{1}+E_{3})}\bar{\rho}_{12}(1)\rho(3)\frac{1}{\rho(3)}F(3)\end{eqnarray*}
where the ellipsis denote additional contributions that can be obtained
by cyclical permutation of the indices $1,2,3$ from those explicitly
displayed inside the square bracket. These three sets of contributions
can be shown to be equal to each other by relabeling the integration
variables: \begin{eqnarray}
 & - & \frac{1}{2}\int\frac{d\theta_{1}}{2\pi}\frac{d\theta_{2}}{2\pi}\mathrm{e}^{-R(2E_{1}+E_{2})}\frac{\bar{\rho}_{13,2}(12)}{\rho(112)}\Big[2\tilde{\rho}(112|\{2,3\})F(12)\nonumber \\
 &  & +2\tilde{\rho}(112|\{1\})F(1)+\tilde{\rho}(112|\{3\})F(2)\Big]\nonumber \\
 & + & \frac{1}{2}\int\frac{d\theta_{1}}{2\pi}\frac{d\theta_{2}}{2\pi}\mathrm{e}^{-R(2E_{1}+E_{2})}\rho(2)\bar{\rho}_{12}(1)\frac{2\tilde{\rho}(11|\{1\})}{\rho(11)}F(1)\nonumber \\
 & + & \frac{1}{2}\int\frac{d\theta_{1}}{2\pi}\frac{d\theta_{2}}{2\pi}\mathrm{e}^{-R(2E_{1}+E_{2})}\bar{\rho}_{12}(1)F(2)\label{eq:dintreshuff}\end{eqnarray}
The terms containing $F(12)$ contribute\begin{equation}
-\int\frac{d\theta_{1}}{2\pi}\frac{d\theta_{2}}{2\pi}F_{4}^{c}(\theta_{1},\theta_{2})\mathrm{e}^{-mR(\cosh\theta_{1}+2\cosh\theta_{2})}\label{eq:dintres1}\end{equation}
where it was used that \[
\frac{\bar{\rho}_{13,2}(12)}{\rho(112)}\tilde{\rho}(112|\{2,3\})=1+O(L^{-1})\]
which results from (\ref{eq:p3ptdeg}) and\[
\tilde{\rho}(112|\{2,3\})=2mL\cosh\theta_{1}+2\varphi(0)+2\varphi(2\theta_{1})++\psi^{(\alpha)}(\theta)+\psi^{(\beta)}(\theta)\]
The terms containing $F(1)$ and $F(2)$ combine to \begin{eqnarray*}
 &  & \int\frac{d\theta_{1}}{2\pi}\frac{d\theta_{2}}{2\pi}\mathrm{e}^{-R(2E_{1}+E_{2})}\left(-\frac{\bar{\rho}_{13,2}(12)}{\rho(112)}\tilde{\rho}(112|\{1\})+\rho(2)\bar{\rho}_{12}(1)\frac{\tilde{\rho}(11|\{1\})}{\rho(11)}\right)F(1)+\\
 &  & \frac{1}{2}\int\frac{d\theta_{1}}{2\pi}\frac{d\theta_{2}}{2\pi}\mathrm{e}^{-R(2E_{1}+E_{2})}\left(-\frac{\bar{\rho}_{13,2}(12)}{\rho(112)}\tilde{\rho}(112|\{3\})+\bar{\rho}_{12}(1)\right)F(2)\end{eqnarray*}
A straightforward (albeit tedious) calculation leads to\begin{eqnarray*}
-\frac{\bar{\rho}_{13,2}(12)}{\rho(112)}\tilde{\rho}(112|\{1\})+\rho(2)\bar{\rho}_{12}(1)\frac{\tilde{\rho}(11|\{1\})}{\rho(11)} & = & -2(\varphi(\theta_{1}-\theta_{2})+\varphi(\theta_{1}+\theta_{2}))+O\left(L^{-1}\right)\\
-\frac{\bar{\rho}_{13,2}(12)}{\rho(112)}\tilde{\rho}(112|\{3\})+\bar{\rho}_{12}(1) & = & -\varphi(\theta_{1}-\theta_{2})-\varphi(\theta_{1}+\theta_{2})+O\left(L^{-1}\right)\end{eqnarray*}
Note that the individual terms in these sums are proportional to $L$
but their contributions drops out. For a more detailed discussion
of such {}``anomalous'' density contributions the reader is referred
to \cite{fftcsa2}.

The total contribution in the $L\rightarrow\infty$ limit turns out
to be just\begin{equation}
-\int\frac{d\theta_{1}}{2\pi}\frac{d\theta_{2}}{2\pi}\mathrm{e}^{-mR(\cosh\theta_{1}+2\cosh\theta_{2})}(2F_{2}^{c}(\theta_{1})+\frac{1}{2}F_{2}^{c}(\theta_{2}))(\varphi(\theta_{1}-\theta_{2})+\varphi(\theta_{1}+\theta_{2}))\label{eq:dintres2}\end{equation}
Summing up the contributions (\ref{eq:res3int}), (\ref{eq:res1int}),
(\ref{eq:dintres1}) and (\ref{eq:dintres2}) the end result is\begin{eqnarray}
\Sigma_{3} & = & \frac{1}{6}\int_{0}^{\infty}\frac{d\theta_{1}}{2\pi}\int_{0}^{\infty}\frac{d\theta_{2}}{2\pi}\int_{0}^{\infty}\frac{d\theta_{3}}{2\pi}\mathrm{e}^{-mR(\cosh\theta_{1}+\cosh\theta_{2}+\cosh\theta_{3})}[F_{6}^{c}(\theta_{1},\theta_{2},\theta_{3})\nonumber \\
 &  & +3F_{4}^{c}(\theta_{2},\theta_{3})(\Phi_{12}+\Phi_{13})+3F_{2}^{c}(\theta_{3})(\Phi_{12}\Phi_{13}+\Phi_{12}\Phi_{23}+\Phi_{13}\Phi_{23})]\nonumber \\
 &  & -\int_{0}^{\infty}\frac{d\theta_{1}}{2\pi}\int_{0}^{\infty}\frac{d\theta_{2}}{2\pi}\mathrm{e}^{-mR(2\cosh\theta_{1}+\cosh\theta_{2})}[F_{4}^{c}(\theta_{1},\theta_{2})+(2F_{2}^{c}(\theta_{1})+\frac{1}{2}F_{2}^{c}(\theta_{2}))\Phi_{12}]\nonumber \\
 &  & +\int_{0}^{\infty}\frac{d\theta_{1}}{2\pi}\mathrm{e}^{-3mR\cosh\theta_{1}}F_{2}^{c}(\theta_{1})\label{eq:our3order}\end{eqnarray}


\begin{thebibliography}{10}
\bibitem{GZ}S. Ghoshal and A.B. Zamolodchikov, \emph{Int. J. Mod.
Phys.} \textbf{A9} (1994) 3841-3886 (Erratum-ibid. \textbf{A9} 4353),
hep-th/9306002.

\bibitem{affleckludwig}I. Affleck and A.W.W. Ludwig, \emph{Phys.
Rev. Lett.} \textbf{67} (1991) 161-164.

\bibitem{friedankonechny}D. Friedan and A. Konechny, \emph{Phys.
Rev. Lett.} \textbf{93} (2004) 030402, hep-th/0312197.

\bibitem{bffprogram}Z. Bajnok, L. Palla and G. Takács, \emph{Nucl.
Phys.} \textbf{B750} (2006) 179-212, hep-th/0603171.

\bibitem{leclairmussardo}A. Leclair and G. Mussardo, \emph{Nucl.
Phys.} \textbf{B552} (1999) 624-642, hep-th/9902075.

\bibitem{fftcsa2}B. Pozsgay and G. Takács, \emph{Nucl. Phys.} \textbf{B788}
(2007) 209-251, arXiv: 0706.3605 {[}hep-th].

\bibitem{bfftcsa}M. Kormos and G. Takács: \texttt{Boundary form factors
in finite volume}, arXiv: 0712.1886 {[}hep-th].

\bibitem{ca2}O.A. Castro-Alvaredo: \texttt{Form factors of boundary
fields for $A_{2}$-affine Toda field theory}, arXiv:0710.0501 {[}hep-th].

\bibitem{BBT}Z. Bajnok, G. Böhm and G. Takács, \emph{J. Phys.} \textbf{A35}
(2002) 9333-9342, hep-th/0207079.\\
 Z. Bajnok, G. Böhm and G. Takács, \emph{Nucl. Phys.} \textbf{B682}
(2004) 585-617, hep-th/0309119. 

\bibitem{bffcount}M. Sz{\H o}ts and G. Takács, \emph{Nucl. Phys.}
\textbf{B785} (2007) 211-233, hep-th/0703226.

\bibitem{bffexp}G. Takács: \texttt{Form factors of boundary exponential
operators in the sinh-Gordon model}, \emph{Nucl. Phys.} \textbf{B},
accepted for publication, arXiv: 0801.0962 {[}hep-th].

\bibitem{Smirnov}F.A. Smirnov: \texttt{Form-factors in completely
integrable models of quantum field theory}, \emph{Adv. Ser. Math.
Phys}. \textbf{14} (1992) 1-208. 

\bibitem{SGff}B. Hou, K. Shi, Y. Wang, W.-l. Yang, \emph{Int. J.
Mod. Phys.} \textbf{A12} (1997) 1711-1741.

\bibitem{lashkevich}M. Lashkevich: \texttt{Boundary form factors
in the Smirnov-Fateev model with a diagonal boundary S matrix}, arXiv:0801.0935
{[}hep-th].

\bibitem{saleurfiniteT}H. Saleur, \emph{Nucl. Phys.} \textbf{B567}
(2000) 602-610, hep-th/9909019.

\bibitem{fftcsa1}B. Pozsgay and G. Takács, \emph{Nucl. Phys.} \textbf{B788}
(2007) 167-208, arXiv: 0706.1445 {[}hep-th].

\bibitem{bbye}P. Fendley and H. Saleur, \emph{Nucl. Phys.} \textbf{B428}
(1994) 681-693, hep-th/9402045.

\bibitem{castrofring}O.A. Castro-Alvaredo and A. Fring, \emph{Nucl.
Phys.} \textbf{B636} (2002) 611-631, hep-th/0203130.

\bibitem{esslerfiniteT}F.H.L. Essler and R.M. Konik: \texttt{Applications
of massive integrable quantum field theories to problems in condensed
matter physics}, cond-mat/0412421. In: Shifman, M. (ed.) et al.: \emph{From
fields to strings}, vol. 1, pp. 684-830. 

\bibitem{tsvelikfiniteTcorr}B.L. Altshuler, R.M. Konik and A.M. Tsvelik,
\emph{Nucl. Phys.} \textbf{B739} (2006) 311-327, cond-mat/0508618.

\bibitem{nondiag}T.R. Klassen and E. Melzer, Nucl. Phys. B382 (1992)
441-485, hep-th/9202034.\\
G. Takacs and G. Watts, Nucl. Phys. B547 (1999) 538-568, hep-th/9810006.

\bibitem{delfinofiniteT}G. Delfino, \emph{J. Phys.} \textbf{A34}
(2001) L161-L168, hep-th/0101180.

\bibitem{mussardodifference}G. Mussardo, \emph{J. Phys.} \textbf{A34}
(2001) 7399-7410, hep-th/0103214.
\end{thebibliography}
\end{document}